\documentclass[aps,prb,preprint]{revtex4-2}
\usepackage{graphicx}
\usepackage{xcolor}
\usepackage{amsmath,amssymb, bm, mathrsfs}
\usepackage{mathtools}
\usepackage{bm}
\usepackage[pdffitwindow,colorlinks,citecolor={red},linkcolor={blue}]{hyperref}
\usepackage[capitalize]{cleveref}
\crefname{section}{Sec.}{Secs.}
\Crefname{section}{Section}{Sections}
\begin{document}

\title{Analytical theory for three wave-mixing processes in a slightly
  deformed cylinder}

\author{Raksha Singla}
\email{raksharuia@icf.unam.mx}
\author{W. Luis Moch\'an}
\email{mochan@icf.unam.mx}

\affiliation{Instituto de Ciencias F\'isicas, Universidad Nacional
  Aut\'onoma de M\'exico, Avenida Universidad s/n, 62210 Cuernavaca,
  Morelos, M\'exico}

\begin{abstract}

The second order optical response of centrosymmetric materials
manifests itself mostly at their surface, being strongly suppressed in
their bulk. However, the overall surface response is also suppressed
in nanoparticles with a centrosymmetric geometry subjected to
homogeneous fields. Nevertheless,
nanoparticles with a {\em noncentrosymmetric geometry} do exhibit second order
optical properties. We develop an
analytical theory to investigate the second order optical response
of a noncentrosymmetric thin cylinder with a slightly
deformed cross-section made up of a centrosymmetric material
subjected to two monochromatic fields. We
calculate the linear and nonlinear near fields perturbatively using the
extent of the deformation away from a circular cross-section as the
perturbation parameter. We obtain expressions for the quadratic
hyperpolarizabilities in terms of the linear response evaluated at the
three frequencies involved. We analyze the
spectral features of the nonlinear response functions and explore their
resonant structure for a model dielectric cylinder. Furthermore, we
evaluate the second order radiated fields, the radiation patterns and
efficiency of the different quadratic processes. We obtain a strong
competition between electric dipolar, magnetic dipolar  and electric
quadrupolar contributions even for very small deformations.
\end{abstract}

\maketitle

\section{Introduction}

Nonlinear optics plays a key role in the development of many modern
photonic technologies such as super continuum generation\cite{dudley},
holography\cite{yariv}, optical parametric amplification\cite{opa},
generation of ultra-short pulses\cite{kaiser, ultra}, etc. Since
optical nonlinear processes are inherently weak, devices made from
conventional nonlinear crystals require strong excitation fields and
phase matching conditions for them to be efficient. Advances in
nanotechnology have led to the development of a plethora of
nanomaterials which exhibit remarkable optical nonlinearities with
unparalleled applications such as in miniaturization of photonic
devices\cite{nanophotonic,rev1nlo}. The nonlinear
dependence on the electric field yields an amplification of the
response in these novel nanomaterials. The presence of processes
that enhance the local field, for example, plasmonic resonances in
metallic nanoparticles have been shown to boost the nonlinear
efficiency\cite{kauranen,czaplicki}. Recently, all
dielectric nanostructures have also been reported with exceptional
nonlinear conversion
efficiencies\cite{diel-nl,thg-diel,gili}. Understanding the underlying
mechanisms of nonlinear processes and their enhancement in these
structures has been a topic of growing interest. Various
nonlinear processes such as second
harmonic generation (SHG)\cite{ulises,raksha}, two photon absorption
(TPA)\cite{tpa}, third harmonic generation (THG)\cite{thg1,thg-diel},
and four wave mixing (FWM)\cite{fwm,fwm1} have been observed in
different nanostructures and their applications have been discussed.

Second order nonlinear processes involve photon-photon
interactions assisted by materials that lead to various three-wave
mixing effects such as the generation of radiation with the sum (SFG)
$(\omega_+=\omega_1+\omega_2)$ or the difference (DFG)
$(\omega_-=\omega_1-\omega_2)$ of the frequencies $\omega_1$ and
$\omega_2$ of two driving fields, conversion of these input
signals to their second harmonics (SHG) $2\omega_1$ or $2\omega_2$,
or the generation of a static quadratic polarization, optical rectification
(OR). The quadratic susceptibility tensor, originating due to
electric-dipolar transitions, is zero within the bulk of a
centrosymmetric medium and hence the second order response manifests
itself mainly at its surface where the inversion symmetry is locally
lost. For this reason, the different second order processes have been
extensively used as surface probes for these class of
materials\cite{sfgimg1,sfgimg2,shgimg1,shgimg2}. Besides providing
remarkable non-invasive surface imaging techniques, they have
demonstrated tremendous potential for numerous other
applications, as in the development of coherent light
sources at different frequencies. For example, SFG has
been used in the production of light sources in the UV-Vis spectral
range\cite{sfggreen,sfg-yellow} and DFG for sources at mid
or far infra-red frequencies\cite{dfg1mir,dfg2mir,dfg3mir}.
Generation of the previously inaccessible terahertz (THz) frequency
band have also been facilitated by
DFG\cite{thzdfg1,thzdfg2}.

The selection rules for the second order
optical properties of centrosymmetric bulk materials are also
applicable to nanoparticles made up of them, with the second order
response being generated largely at their surfaces.
However, for particles with a centrosymmetric shape, an exact
cancellation of the induced quadratic polarization from opposite
points of the surface leads to a null overall response. A second order response,
may still be observed in such cases, but it is due to multipolar
excitations. On the other hand, dipole driven second order response
from particles with
{\em noncentrosymmetric geometry} may be observed, as the local
contributions from each point of the surface do not
cancel. SHG\cite{bachelier,zhou,czaplicki} and
DFG\cite{ciraci-dfg,fdtddf} from noncentrosymmetric nanoparticles or
nanostructures have been studied extensively, using both experimental
and numerical means.

In a previous work we have developed an
analytical theory for the optical SHG of a
noncentrosymmetric cylinder\cite{raksha}. Here, we present a
calculation of the response using an approximate
analytical perturbative theory and generalize it to explore
all the second order optical processes, namely, SFG, DFG, OR, and the
previously discussed SHG, though including in the latter the
possibility of excitation by noncollinear fundamental
fields. To this end, we choose an isolated long cylinder with a
noncentrosymmetric cross-section, as in Ref. \cite{raksha}.
We
consider two monochromatic fields with polarizations
normal to the axis of the cylinder, and compute
the linear and the nonlinear
fields induced within and outside the cylinder, generalizing
the perturbative approach of
Ref. \cite{raksha}. We employ the {\em dipolium} model\cite{dipolium}
to obtain the nonlinear
response within the bulk and on the surface of the cylinder. That
model was originally developed to explore the SH
response,  and was later
extended towards the SF response of conductors\cite{hydro}. Within the dipolium
theory, the material is modeled
as a homogeneous array of polarizable entities that
respond harmonically to the electromagnetic field. The origin of the
nonlinearity in the model is the non-homogeneity of the fields,
including their abrupt variation
across  interfaces. For the sake of
completeness, we present a brief description of the model and a
derivation of the DF nonlinear surface and bulk susceptibilities which we
write in terms of some dimensionless parameters\cite{rudnick} that
depend on the linear dielectric response evaluated at the frequencies
involved. Then we extend our results to get also the SF, SH and OR
responses. Furthermore, we calculate the nonlinear fields in
the radiation zone and analyze the efficiency of the different second
order processes.

The structure of the paper is the following. In \cref{dfg}, we
describe our theory to calculate the generation of a DF signal from a
planar surface. In \cref{nanowire} we calculate the DF
response of the cylinder, assuming it is locally flat. We find the
DF efficiency in \cref{dfconveff} and generalize it to
the SF, SH and OR cases. \Cref{results} illustrates our
results for a model dielectric cylinder. Finally, we present our conclusions
in \cref{conclusions}.


\section{Theory}\label{th}

\subsection{Response of a semi-infinite system}\label{dfg}

We consider a  semi-infinite {\em
  dipolium} \cite{dipolium}, a homogeneous array of harmonic
polarizable entities. Each
polarizable entity is represented by an electron of charge
$-e$ and mass $m$ at a separation $\bm{x}$ from its equilibrium position
$\bm{r}_0$ to which it is bound by a harmonic force with resonant
frequency $\omega_0$. Its classical equation of motion under the
influence of a spatially varying external electromagnetic field is
\begin{equation}\label{eqnmotion}
  m\ddot{\bm{x}}=-m\omega_0^2\bm{x}
  -\frac{m}{\tau}\dot{\bm{x}}-e\bm{E}(\bm{r},t)
  -\frac{e}{c}\dot{\bm{x}}\times\bm{B}(\bm{r},t),
\end{equation}
where we have included a dissipative term characterized by a lifetime
$\tau$. We remark that the fields should be evaluated at the actual
position $\bm r =\bm{r}_0+\bm{x}$ of the electron, not at its
equilibrium position $\bm r_0$. Assuming the displacement
$\bm{x}$ from the equilibrium position to be smaller than the scale of variation in
the driving fields, we perform a Taylor expansion
\begin{equation}\label{taylor}
  \bm{E}(\bm{r}_0+\bm{x},t)\approx\bm{E}(\bm{r}_0,t)
  +\bm{x}\cdot\nabla\bm{E}(\bm{r}_0,t)+\ldots
\end{equation}
Substituting \cref{taylor} in the equation of motion
(\ref{eqnmotion}), we get
\begin{equation}\label{eqnmotion1}
  m\ddot{\bm{x}}=-m\omega_0^2\bm{x}
  -\frac{m}{\tau}\dot{\bm{x}}-e\bm{E}(\bm{r}_0,t)
  -e\,\bm{x}\cdot\nabla\bm{E}(\bm{r}_0,t)
  +e\,\dot{\bm{x}}\times
  \int^{t}_{-\infty}dt'\,\nabla\times\bm{E}(\bm{r_0},t')+\ldots,
\end{equation}
where we have written the magnetic field $\bm{B}(\bm{r},t)
=-c\int^{t}_{-\infty}dt'\,\nabla\times\bm{E}(\bm{r},t')$ in terms of
the electric field, and we assume the electromagnetic field is switched on
adiabatically.  Notice that the coefficients of
$\bm x$ and $\dot{\bm x}$ in the last two terms depend on time through
the spatial derivatives of the field. Thus, \cref{eqnmotion1} is the
equation of a forced, damped harmonic oscillator whose effective {\em stiffness}
varies in time, making
it similar to a parametric oscillator.
Hence, even though the harmonic oscillator is considered the paradigmatic
linear system,
it becomes nonlinear through the spatial variations of the driving fields.

We now drive the system with two  electromagnetic waves oscillating at frequencies
$\omega_1$ and $\omega_2$
\begin{equation}\label{ext}
  \bm E(\bm r,t)= \bm E_1(\bm r )e^{-i\omega_1t}
  + \bm E_2(\bm r )e^{-i\omega_2t}+ c.c.,
\end{equation}
where $\bm E_1(\bm r )$ and $\bm E_2(\bm r )$
are complex amplitudes and $c.c.$ stands
for the complex conjugate. Since the incident optical fields are
usually much smaller than the microscopic atomic fields, we employ a
perturbative approach to solve \cref{eqnmotion1} by expanding the
solution in powers of $\bm{E}$,
\begin{equation}\label{try}
  \bm{x}(t)=\bm{x}^{(1)}(t)+\bm{x}^{(2)}(t)+\cdots.
\end{equation}

The linear solution
$\bm{x}^{(1)}(t)=\sum_g\,\bm{x}_g^{(1)}e^{-i\omega_gt}+c.c.$,
with $g=1,2$ is a superposition of two oscillations
with amplitudes $\bm{x}^{(1)}_g$ corresponding to
the incident frequencies $\omega_g$ respectively,
each obeying the equation of a forced linear harmonic oscillator
\begin{equation}\label{linear}
  -m\omega_g^2\bm{x}_g^{(1)}= -m\omega_0^2\bm{x}_g^{(1)}
  +im\frac{\omega_g}{\tau}\bm{x}_g^{(1)}
  -e\bm{E}_g(\bm{r}_0,t),
\end{equation}
whose solution yields the
induced linear dipole moment
$\bm{p}_g^{(1)}=-e\bm{x}_g^{(1)}=\alpha_g\bm{E}_g$ where
each $\alpha_g=\alpha(\omega_g)$ is the linear polarizability
\begin{equation}\label{alphalin}
  \alpha(\omega)=\frac{e^{2}/m}{\mathscr{D}(\omega)},
\end{equation}
evaluated at frequency $\omega_g$, and
\begin{equation}
  \label{eq:Denominator}
\mathscr{D}(\omega)=\omega_0^2-\omega^2-i\omega/\tau.
\end{equation}

We employ the abbreviated notation $f_{g}\equiv
f(\omega_g)$ for any function $f$ dependent on frequency. Now we write
the second order contribution to \cref{eqnmotion1},
\begin{align}\label{2ndorder}
  m\ddot{\bm{x}}^{(2)}(t)=& -m\omega_0^2\bm{x}^{(2)}(t)
  -\frac{m}{\tau}\dot{\bm{x}}^{(2)}(t)
  -e\bm{x}^{(1)}(t)\cdot\nabla\bm{E}(\bm{r}_0,t)\nonumber\\
     &+e\dot{\bm{x}}^{(1)}(t)\times\int^{t}dt'\nabla\times
  \bm{E}(\bm{r_0},t').
\end{align}
Notice that upon substitution of $\bm {x}^{(1)}$, the driving terms in
\cref{2ndorder}
become quadratic in $\bm{E}$ with several frequency components: DC,
the second harmonic of both incident
frequencies $2\omega_g$, the sum frequency
$(\omega_{+}\equiv\omega_1+\omega_2)$, and the difference frequency
$(\omega_{-}\equiv\omega_1-\omega_2)$.  The
equation corresponding to DF is
\begin{align}\label{2nd}
   \omega_{-}^2\,{\bm{x}}^{(2)}_-&=\omega_0^2\bm{x}^{(2)}_-
  \,-\,i\frac{\omega_{-}}{\tau}{\bm{x}}^{(2)}_-
  +\frac{e}{m}\left(\bm{x}^{(1)}_1\cdot\nabla\bm{E}_2^*+
    \bm{x}^{(1)*}_2\cdot\nabla\bm{E}_1\right)\nonumber\\
 &+\frac{e}{m}\left[\left(\frac{\omega_1}{\omega_2}\right)\bm{x}^{(1)}_1
    \times\nabla\times \bm{E}_2^*
   +\left(\frac{\omega_2}{\omega_1}\right)\bm{x}^{(1)*}_2
   \times\nabla\times \bm{E}_1\right],
\end{align}
where  the superscript $(*)$ on any quantity denotes its complex
conjugate and the subscript ${-}$ means the corresponding terms are
evaluated at $\omega_-$. We solve \cref{2nd} to obtain
the quadratic DF dipole moment
${\bm p}^{(2)}_-=-e{\bm x}^{(2)}_-$,
\begin{align}\label{DFdip}
  {\bm p}^{(2)}_-=
  -\frac{1}{e}\alpha_-
  \Bigg[&\alpha_1
    \left({\bm E}_1\cdot\nabla{\bm E}^*_2+\frac{\omega_1}{\omega_2}
      {\bm E}_1\times(\nabla\times{\bm E}^*_2)\right)\nonumber\\
    &+\alpha^*_2
    \left({\bm E}^*_2\cdot\nabla{\bm E}_1+\frac{\omega_2}{\omega_1}
         {\bm E}^*_2\times(\nabla\times{\bm E}_1)\right)\Bigg].
\end{align}
There are two other second order moments \cite{hydro}
which contribute
to the nonlinear DF response: the electric
quadrupole moment and the magnetic dipole moment. For convenience, we
define the quadratic
electric quadrupole moment as ${\bm Q}^{(2)}_-=-\mathrm{e}{\bm x}^{(1)}_1{\bm x}^{(1)*}_2
-\mathrm{e}{\bm x}^{(1)*}_2{\bm x}^{(1)}_1.$
This differs from the usual definition, which is
{\em traceless} and includes a
numerical prefactor of $3$. Similarly, the DF magnetic moment is given
by ${\bm \mu}^{(2)}_-= (-e/2mc)\{{\bm x}^{(1)}_1\times m\dot{\bm
  x}^{(1)*}_2 +{\bm x}^{(1)*}_2\times m\dot{\bm x}^{(1)}_1\},$
From the linear
solution we obtain
\begin{equation}\label{DFquad}
 {\bm Q}^{(2)}_-=-\frac{1}{e}\alpha_1\alpha^*_2
  ({\bm E}_1{\bm E}^*_2+{\bm E}_2^*{\bm E}_1).
\end{equation}
and
\begin{equation}\label{DFmag}
  {\bm \mu}^{(2)}_-=
  -\frac{i}{2ce}\alpha_1\alpha^*_2\,(\omega_1+\omega_2)\,
  ({\bm E}_1\times{\bm E}^*_2).
\end{equation}
We must remark here that although these nonlinear moments have
been calculated through a classical model, expressions equivalent to
the above are obtained for a quantum harmonic
oscillator which interacts with the perturbing electromagnetic field
through electric-dipolar, magnetic-dipolar and electric-quadrupolar
transitions\cite{recamier}.

We consider now a semi-infinite system made from $n$ of these
polarizable entities
per unit volume. We assume the system is translationally invariant
along the $x-y$ plane and that its surface lies at $z=0$, across which
$n(z)$ changes rapidly albeit continuously from its
bulk value $n(z\to\infty)=n_B$ to its vacuum value
$n(z\to-\infty)=0$.
The macroscopic nonlinear polarization induced in
the system is
\begin{equation}\label{macronlp}
  {\bm P}_{-,\text{src}}(z)=n(z){\bm
    p}^{(2)}_-
  - \frac{1}{2}\nabla\cdot\left(n(z){\bm Q}^{(2)}_-\right)
  +\frac{ic}{\omega_{-}}\nabla\times\left(n(z) {\bm
      \mu^{(2)}_-}\right).
\end{equation}
Note that the above expression has the usual electric dipole moment
density and an additional term related the non-homogeneity of the
electric quadrupolar moment density\cite{jackson}. Furthermore, it contains a
term related to the curl of the quadratic magnetic moment
density. This term is not conventional, but it yields the same induced
current $\bm j_-=\partial \bm P_-/\partial t$ and is more convenient
than keeping only the first two terms
in \cref{macronlp} and adding a nonlinear
magnetization\cite{agranovich1} and the corresponding magnetization
current. The nonlinear polarization
(\ref{macronlp}) is a nonlinear source oscillating at the difference
frequency, and it generates an oscillating DF field $\bm E_-$. The linear
response of the system to this DF field yields an additional DF
polarization, so that substituting \cref{DFdip,DFquad,DFmag} into
\cref{macronlp} we get
the screened self-consistent nonlinear polarization
\begin{align}\label{macroPnl}
  {\bm P}_{-}(z)=&
                n(z)\alpha_{-}{\bm E}_{-}
              -\frac{n(z)}{e}\alpha_-\Bigg[\alpha_1
    \left({\bm E}_1\cdot\nabla{\bm E}^*_2+\frac{\omega_1}{\omega_2}
      {\bm E}_1\times(\nabla\times{\bm E}^*_2)\right)\nonumber\\
    &+\alpha^*_2
    \left({\bm E}^*_2\cdot\nabla{\bm E}_1+\frac{\omega_2}{\omega_1}
         {\bm E}^*_2\times(\nabla\times{\bm E}_1)\right)\Bigg]\nonumber\\
     ~& +\frac{1}{2e}\alpha_1\alpha^*_2\nabla\cdot n(z)
         ({\bm E}_1{\bm E}^*_2+{\bm E}_2^*{\bm E}_1)\nonumber\\
              ~& +\frac{1}{2e}
                 \alpha_1\alpha^*_2\left(
                 \frac{\omega_1+\omega_2}{\omega_{-}}\right)
                  \nabla(\times n(z) ({\bm E}_1\times{\bm E}^*_2)).
\end{align}

In order to find the {\em surface response} of the medium, we will only be
interested in the thin {\em selvedge} region whose thickness we can
assume is much
smaller than the wavelength, allowing us to safely ignore within
it the
effects of retardation. Thus, we identify $\bm E_-$ with the
depolarization field
\begin{equation}\label{Dep}
  E_{-,i}=-4\pi P_{-,z}\delta_{iz},
\end{equation}
we drop the $\nabla\times\bm E_g$ terms and we ignore the relatively
slow spatial variations of the field along the surface. The surface
polarization can be obtained after solving \cref{macroPnl} for $\bm
P_-$ and integrating it across the selvedge,
\begin{equation}\label{Psnormal}
  {\bm P}^s_-=\int_{\text{se}}dz\,{\bm P}_{-}(z),
\end{equation}
where $\text{se}$ denotes the selvedge. We define the components of
the DF
quadratic surface susceptibility tensor through
\begin{equation}\label{chi-general}
  P^s_{-,i}=\sum_{jk}
  \left[\chi^s_{ijk}(\omega_1,\overline{\omega}_2)
  +
  \chi^s_{ikj}(\overline{\omega}_2,\omega_1)\right] F_{1,j}F^*_{2,k},
\end{equation}
where $i,j,k$ denote Cartesian
components and $\overline{\omega}_2$ denotes $-\omega_2$. Here, $\bm
F_g=(E_{g,x}, E_{g,y}, D_{g,z})$
is a
field whose components are the corresponding components of either $\bm
D_g$ or $\bm E_g$ that are continuous across the surface. We use $\bm F_g$
to avoid the ambiguity
about the position in the selvedge where the fields are to be
calculated. Note that $j$ and $k$ are dummy indices and thus can be
interchanged. Thus, we may impose the
intrinsic permutation symmetry
$\chi^s_{ijk}(\omega_1,\overline{\omega}_2)
=\chi^s_{ikj}(\overline{\omega}_2,\omega_1)$.

From \cref{macroPnl,Dep} we obtain the normal component of the
macroscopic polarization in the selvedge region,
\begin{align}
  P_{-,z}(z)=&\frac{1}{e \epsilon_{-}(z)}
            \Bigg[-n(z)\alpha_{-}\left(\alpha_1
                                  \frac{1}{\epsilon_1(z)}
              \frac{\partial}{\partial z}\frac{1}{\epsilon_2^*(z)}
              +\alpha^*_2\frac{1}{\epsilon_2^*(z)}
 \frac{\partial}{\partial z} \frac{1}{\epsilon_1(z)}\right)\nonumber\\
 &+\alpha_1\alpha^*_2\frac{\partial}{\partial z}
              n(z)\frac{1}{\epsilon_1(z)}\frac{1}{\epsilon_2^*(z)}
              \Bigg]D_{1,z}D^*_{2,z}+1 \leftrightarrow 2,\label{Pnlnormal}
\end{align}
where we
introduced the permittivity
\begin{equation}\label{epsilon}
  \epsilon_g(z)=1+4\pi n(z)\alpha_g,
\end{equation}
used $E_{g,z}(z)=D_{g,z}/\epsilon_g(z)$ and assumed $D_{g,z}$
is constant across the selvedge for $g=1,2,-$.

The resulting normal component of the DF polarization depends on
$z$ through the density profile $n(z)$ and its spatial
derivatives and is large only within the thin
selvedge region where the linear response has a large
gradient, and in our long wavelength approximation it vanishes in the
bulk and in vacuum. We now integrate \cref{Pnlnormal} by substituting
it in \cref{Psnormal}. The integral to be evaluated is of the form
$\int dz f(n(z))dg(n(z))/dz$, where $f$ and $g$ are rational functions
of $n(z)$. The integration can be divided into different intervals
where $n$ varies monotonically, which allows us to change integration
variable $z \to n$. As the
integrands are rational functions of $n$, they may be
integrated analytically for any density profile $n(z)$ to obtain the
normal component of the
nonlinear surface polarization
\begin{equation}\label{Psperp}
  {P}^s_{-,z}=\chi^s_{zzz}(\omega_1,\overline{\omega}_2)
    D_{1,z}D^*_{2,z}+1\leftrightarrow 2,
\end{equation}
where
\begin{equation}\label{chiperpDF}
  \chi^s_{zzz}(\omega_1,\overline{\omega}_2)
  =-\frac{a(\omega_1,\overline{\omega}_2)}{64 \pi^2 n_B e}
  \left(\frac{\epsilon_1-1}{\epsilon_1}\right)
  \left(\frac{\epsilon^*_2-1}{\epsilon^*_2}\right),
\end{equation}
and
\begin{equation}\label{aDF}
  a(\omega_1,\overline{\omega}_2)=-2\left[1+
    \frac{\big(1-\epsilon_-\big)\,
      \epsilon_1\epsilon^*_2\,
      \big(\epsilon^*_2
      \log(\epsilon_-/\epsilon_1)
      +{\mathrm{c.p.}}\big)}
   {\big(\epsilon_1-\epsilon^*_2\big)\,\big(\epsilon^*_2
      -\epsilon_-\big)\,
           \big(\epsilon_--\epsilon_1\big)}\right]
\end{equation}
is a dimensionless quantity that
parameterizes the normal component of the nonlinear surface
polarization\cite{rudnick}. Here, c.p. denotes the terms obtained from
the previous one through cyclic permutations of the three indices $1,2,-$.

\begin{figure}
\hspace{-0.5cm}
  \begin{tabular}{c c}
   \includegraphics[width=0.5\linewidth,trim={0 0 0 1cm},clip]
    {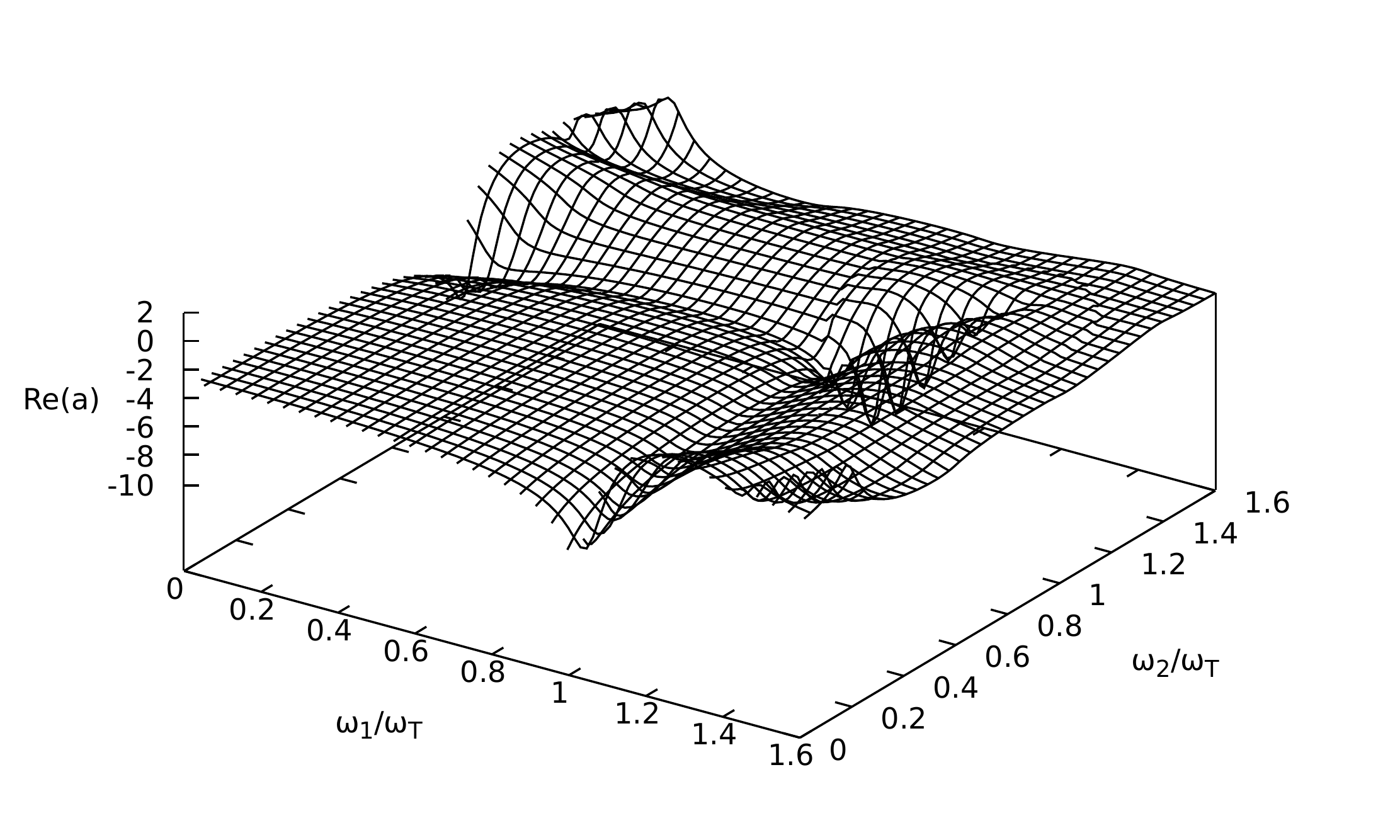}&
    \includegraphics[width=0.5\linewidth,trim={0 0 0 1cm},clip]
    {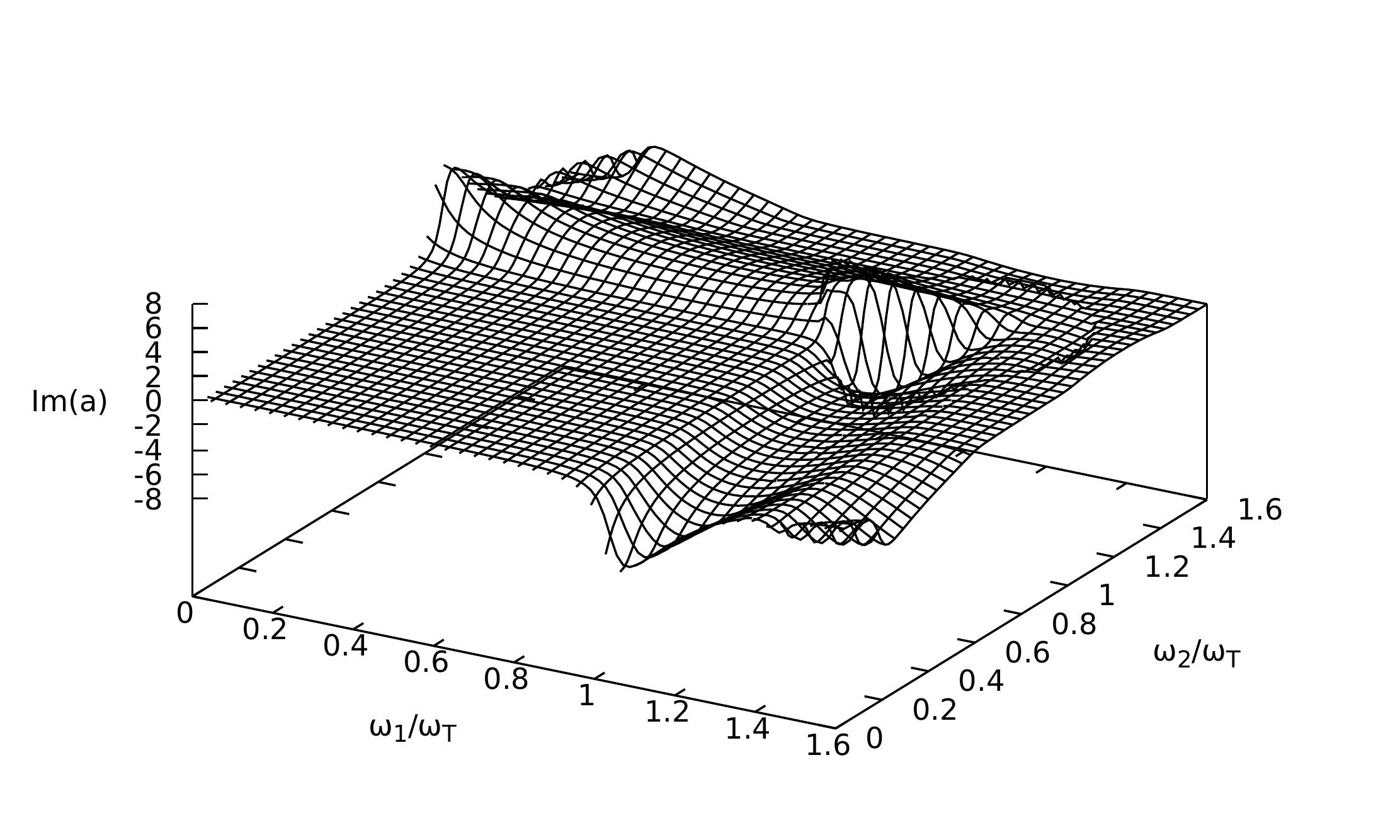}
  \end{tabular}
  \caption{Real (left panel) and imaginary (right  panel) part of
    $a(\omega_1,\overline{\omega}_2)$ for a harmonic solid (\cref{gendiel})
    with $\omega_{\mathrm{L}}=\sqrt 2\,\omega_{\mathrm{T}}$ and
    $\tau=20/\omega_{\mathrm{T}}$, as function of $\omega_1$ and $\omega_2$.}
  \label{reim-a-df}
\end{figure}

In \cref{reim-a-df} we illustrate the behavior of the real and
imaginary parts of $a(\omega_1,\overline{\omega_2})$ for a model
harmonic solid whose dielectric function\cite{ashcroft}
\begin{equation}\label{gendiel}
\epsilon_{\mathrm{d}}(\omega)
=\frac{\omega_{\mathrm{L}}^2-\omega^2-i\omega/\tau}
  {\omega_{\mathrm{T}}^2-\omega^2-i\omega/\tau}
\end{equation}
has a
single Lorentzian resonance, where $\omega_{\mathrm{L}}$ and
$\omega_{\mathrm{T}}$ are the
frequencies of
the longitudinal and transverse optical modes respectively and we
included a small dissipation characterized by $\tau$. We choose
$\omega^2_{\mathrm{L}}=2\omega^2_{\mathrm{T}}$ and
$\tau=20/\omega_{\mathrm{T}}$. Between its pole at
$\omega_{\mathrm{T}}$ and its zero at $\omega_{\mathrm{L}}$, the
dielectric function is negative and
the logarithm in \cref{aDF} becomes large. Hence, we expect the real
and imaginary part of $a$ to exhibit spectral features in this
region. \cref{reim-a-df} shows peaks and valleys for both the real
and imaginary parts of $a$ whenever $\omega_1$ or $\omega_2$
falls in this region. Moreover, a broad valley along the diagonal
$\omega_1\approx\omega_2$ is observed in the region where both input
frequencies lie in this region, with their difference frequency close
to zero. The parameter $a$ has a constant value for low frequencies
and reaches its asymptotic value \cite{petukhov3} of $-2$ for high frequencies.

We follow a similar procedure to that shown above for the parallel
component of the nonlinear polarization. Using \cref{macroPnl}, we
find
\begin{align}
  {\bm P}_{-,\parallel}(z)
  =\frac{1}{2e}\alpha_1\alpha^*_2\Bigg[&
  {\bm E}_{1,\parallel}\frac{\partial}{\partial z}
  \frac{n(z)}{\epsilon_2^*(z)}D_{2,z}^*+{\bm E}_{2,\parallel}^*
  \frac{\partial}{\partial z}\frac{n(z)}{\epsilon_1(z)}D_{1,z}\nonumber\\
  &+\left(\frac{\omega_1+\omega_2}{\omega_{-}}\right)\left(
    \frac{\partial}{\partial z}\frac{n(z)}{\epsilon_2^*(z)}
    D_{2,z}^*{\bm E}_{1,\parallel}-\frac{\partial}{\partial z}
    \frac{n(z)}{\epsilon_1(z)}D_{1,z}{\bm E}_{2,\parallel}^*
  \right)\Bigg],\label{Pnlpar}
\end{align}
which we integrate across the selvedge to obtain
the nonlinear tangential surface
polarization
\begin{align}\label{Pspar}
  \bm{P}^s_{-,\parallel} &=\int_{-\infty}^{\infty} dz {\bm
    P}_{-,\parallel}(z)
    \nonumber \\
  &=\frac{1}{2e}\alpha_1\alpha^*_2
    \left[\frac{n_B}{\epsilon^*_2}
    \left(1+\frac{\omega_1+\omega_2}{\omega_{-}}\right)
    \bm{E}_{1,\parallel}D^*_{2,z}
    +\frac{n_B}{\epsilon_1}
    \left(1-\frac{\omega_1+\omega_2}{\omega_{-}}\right)
D_{1,z}\bm{E}^*_{2,\parallel}\right]\nonumber\\
&=\chi^s_{\parallel\parallel z}
  (\omega_1,\overline{\omega}_2)\bm{E}_{1,\parallel}D^*_{2,z}
  +\chi^s_{\parallel z\parallel}
  (\omega_1,\overline{\omega}_2)D_{1,z}\bm{E}^*_{2,\parallel}
  +1\leftrightarrow 2,
\end{align}
where the surface susceptibility is parameterized as
\begin{equation}\label{chipar1}
  \chi^s_{\parallel\parallel z}(\omega_1,\overline{\omega}_2)
  =\chi^s_{\parallel z\parallel}(\overline{\omega}_2,\omega_1)
  =-\frac{1}{32\pi^2 n_B e}
  \frac{(\epsilon_1-1)(\epsilon^*_2-1)}{\epsilon^*_2}
    \frac{\omega_1}{\omega_{-}}b(\omega_1,\overline{\omega}_2),
\end{equation}
\begin{equation}\label{chipar2}
  \chi^s_{\parallel\parallel z}(\overline{\omega}_2,\omega_1)
  =\chi^s_{\parallel z\parallel}(\omega_1,\overline{\omega}_2)
  =\frac{1}{32\pi^2 n_B e}
  \frac{(\epsilon_1-1)(\epsilon^*_2-1)}{\epsilon_1}
    \frac{\omega_2}{\omega_{-}}b(\overline{\omega}_2,\omega_1),
\end{equation}
with
\begin{equation}\label{b}
  b(\omega_1,\overline{\omega}_2)=b(\overline{\omega}_2,\omega_1)=-1.
\end{equation}
There is another component $\chi^s_{z\|\|}$  of the surface
susceptibility tensor allowed by the in-plane isotropy of the
surface\cite{rudnick}, but it is null within our model.

We now focus on the bulk quadratic polarization of the system, which
we find by substituting \cref{DFdip,DFquad,DFmag} in
\cref{macronlp},
\begin{equation}\label{bulkDF}
  \bm{P}_{-,\text{src}}^{B}=D_1\bm{E}^*_2\cdot(\nabla\bm{E}_1)
  +\tilde{D}_1(\bm{E}^*_2\cdot\nabla)\bm{E}_1+ 1\leftrightarrow 2,
\end{equation}
where
\begin{align}
  D_g&=\frac{1}{16 \pi^2 n_{B}e}(\epsilon^*_2-1)(\epsilon_1-1)\delta_g d_g,
  \label{DFbulkD}\\
  \tilde{D}_g&=\frac{1}{16 \pi^2 n_{B}e}(\epsilon^*_2-1)
  (\epsilon_1-1)\tilde{\delta}_g \tilde{d}_g,\label{DFbulkDbar}
\end{align}
with
\begin{align}
  \delta_1&=-\left(\frac{\epsilon_{-}-1}{\epsilon_1-1}\right)
            \frac{\omega_2}{\omega_1},\label{deltadf1}\\
  \tilde{\delta}_1&=\frac{\omega_1}{\omega_{-}}
                    -\left(\frac{\epsilon_{-}-1}{\epsilon_1-1}\right)
                    \frac{\omega_{-}}{\omega_1},
                       \label{deltabardf1}
\end{align}
\begin{align}
  \delta_2&=-\left(\frac{\epsilon_{-}-1}{\epsilon^*_2-1}\right)
            \frac{\omega_1}{\omega_2},\label{deltadf2}\\
  \tilde{\delta}_2&=-\frac{\omega_2}{\omega_{-}}
                    +\left(\frac{\epsilon_{-}-1}{\epsilon^*_2-1}\right)
                    \frac{\omega_{-}}{\omega_2},
                       \label{deltabardf2}
\end{align}
and $d_1=d_2=\tilde{d}_1=\tilde{d}_2=1$. Note that $P^B_{-,\text{src}}$
plays the role of an external source for the DF
field in the bulk. The total bulk polarization contains also
the polarization linearly induced in response to the
self-consistent DF field, as shown in \cref{macroPnl}. To obtain the
self-consistent DF field Maxwell's equations with sources should be
solved with appropriate boundary conditions.

In the dipolium model we
assumed all entities to be identical to each other, so it doesn't
account for effects such
as those arising from the surface electronic structure. We must
remark that in a
real system these additional effects may not be negligible and must be
accounted for in more realistic models. Here,
we only focused on the effect of the spatial variation of the field
on the second order response.

By construction, the dipolium
model above corresponds to a dielectric material. However, it
may be shown that its results are equivalent to those of the
local jellium model\cite{hydro}, so they may be applied to metals
simply by substituting their corresponding dielectric functions. We
recall that the results of the dipolium model are valid even for a
quantum harmonic oscillator interacting with a perturbing
electromagnetic field through electric dipole, electric quadrupole,
and magnetic dipole transitions\cite{recamier}.

%
\subsection{Response of  an isolated nanocylinder}\label{nanowire}

We will now consider an isolated, long cylinder with a
noncentrosymmetric geometry but made up of a centrosymmetric
material with a nanometric radius. We assume translational symmetry along the axis of the
cylinder ($\hat{z}$ direction), disregarding edge effects as if the
cylinder were infinitely long, allowing us to perform all
calculations in $2D$. We consider a cross-section slightly deformed away
from a symmetric circle, described in polar coordinates by
\begin{equation}\label{rtheta}
  r_s(\theta)=r_{0}(1+\xi\cos3\theta),
\end{equation}
where $r_0$ is the radius of the nominal cylinder and $\xi$
characterizes the extent of deformation.
This is the most simple
noncentrosymmetric deformation of a circle, consisting of three lobes
an angle of $2\pi/3$ apart.
The SH for this shape was studied in
Ref. \cite{raksha}.

We first excite the system with two
monochromatic fields oscillating at frequencies $\omega_1$ and
$\omega_2$ with polarization on the plane of the
cross-section and perpendicular to each other,
\begin{equation}\label{dfnwEext}
  \bm E^{\mathrm{ext}}(t)=
  E_1e^{-i\omega_1 t}\hat{\bm x}+
  E_2e^{-i\omega_2 t}\hat{\bm y}+c.c.,
\end{equation}
where $E_1$ and $E_2$ are complex amplitudes and we take the
corresponding polarization along $\hat{\bm x}$ and $\hat{\bm y}$
respectively. We
disregard the spatial dependence of the fields to concentrate on the
effects of the noncentrosymmetric geometry, which is consistent with
a long wavelength approximation $r_0\ll\lambda_g$ where $\lambda_g$
($g=1,2$) are the wavelengths of the incoming waves. We
follow the perturbative approach introduced in Ref. \cite{raksha}
to evaluate the self-consistent induced near fields. We start with the
general nonretarded solution
$\phi_g(r,\theta) = \phi^{\mathrm{in}}_{g}(r,\theta)
\Theta(r_s(\theta)-r) + \phi^{\mathrm{out}}_{g}(r,\theta)
\Theta(r-r_s(\theta))$
in polar coordinates ($r,\theta$) of Laplace's equation for the scalar
potential within the particle and in its neighborhood,
\begin{subequations}\label{phitrial}
  \begin{align}
   \phi^{\mathrm{in}}_{g}(r, \theta) &= \sum_{l=0}^{\infty} r^{l}
                                       (s_{gl}\cos l\theta+t_{gl} \sin l\theta),
                                       \label{phiin}
                            \\
  \phi^{\mathrm{out}}_{g}(r,\theta) &= \phi_g^{\mathrm{ex}}
                            + \sum_{l=0}^{\infty}r^{-l}
                            (u_{gl}\cos l\theta+v_{gl}\sin l
                            \theta),
                            \label{phiout}
  \end{align}
  \end{subequations}
where $\phi^{\mathrm{ex}}_1(r,\theta)=-E_1r\cos\theta$,
$\phi^{\mathrm{ex}}_2(r,\theta)=-E_2r\sin\theta$ and $\Theta$ is the unit step
function. We expand
the multipolar coefficients $\zeta_{gl}$  (any of
$s_{gl}$, $t_{gl}$, $u_{gl}$ or $v_{gl}$) as power series on the
deformation parameter $\xi$, $\zeta_{gl}=\sum_{n=0}^\infty \zeta_{gl}^{(n)}
\xi^n$. As mentioned previously, we
restrict ourselves to small deformations, and we consider terms up to
linear order in $\xi$ only.  Using
Eqs. (\ref{phitrial}) and imposing boundary conditions \cite{jackson} at the
interface $r=r_s(\theta)$, we obtain the self-consistent linear
potential,
\begin{subequations}\label{dfnwphi12}
\begin{align}
  \frac{\phi^{\mathrm{out}}_1}{E_1}&=-r\cos\theta -
\frac{1-\epsilon_1}{1+\epsilon_1}
\frac{r_{0}^2}{r}\cos\theta  + \xi\left[\left
(\frac{1-\epsilon_1}{1+\epsilon_1}
  \right)^2 \frac{r_{0}^3}{r^2}\cos2\theta
  -\frac{1-\epsilon_1}{1+\epsilon_1}
  \frac{r_{0}^5}{r^4}\cos 4\theta\right],
  \label{dfnwphi1out}\\
  \frac{\phi^{\mathrm{in}}_1}{E_1}&=
             -\frac{2}{1+\epsilon_1}r\cos\theta +
  2\xi
\frac{1-\epsilon_1}{(1+\epsilon_1)^2}
\frac{r^2}{r_{0}} \cos 2\theta,
  \label{dfnwphi1in}\\
  \frac{\phi^{\mathrm{out}}_2}{E_2}&=-r\sin\theta -
  \frac{1-\epsilon_2}{1+\epsilon_2}
\frac{r_{0}^2}{r}\sin\theta  - \xi\left[
\left(\frac{1-\epsilon_2}{1+\epsilon_2}
  \right)^2 \frac{r_{0}^3}{r^2}\sin2\theta
  +\frac{1-\epsilon_2}{1+\epsilon_2}
  \frac{r_{0}^5}{r^4}\sin 4\theta\right],
  \label{dfnwphi2out}\\
  \frac{\phi^{\mathrm{in}}_2}{E_2}&=
-\frac{2}{1+\epsilon_2}r\sin\theta -
2\xi\frac{1-\epsilon_2}{(1+\epsilon_2)^2}
\frac{r^2}{r_{0}} \sin 2\theta.
  \label{dfnwphi2in}
\end{align}
\end{subequations}

The spatial variations of the self-consistent  linear fields
$\bm E_{g}=-\nabla\phi_{g}$, induce a macroscopic nonlinear
polarization within the bulk of the cylinder given by \cref{macronlp}
but with a position independent density $n=n_B$,
\begin{align}
  \bm P_{-,\text{src}}=&\frac{E_1E_2^*\xi}{2\pi^2ne r_0}
\frac{(1-\epsilon_{1})(1-\epsilon_2^*)}
                         {(1+\epsilon_1)^2(1+\epsilon_2^*)^2}
     \Bigg[-(1-\epsilon_{-})(2+\epsilon_1
                                +\epsilon_2^*)
                +(1-\epsilon_1\epsilon_2^*)\nonumber\\
  &+\left(\frac{\omega_1+\omega_2}{\omega_{-}}\right)
    (\epsilon_2^*-\epsilon_1)
    \Bigg]\left\{\sin\theta\,\hat{r}+\cos\theta\,\hat{\theta}\right\}.
    \label{dfnwPnl}
\end{align}

The nonlinear bulk polarization induces
a null charge density within the cylinder,
\begin{equation}\label{dfnwrho}
\rho_{-,\text{src}}=-\nabla\cdot\bm P_{-,\text{src}}=0,
\end{equation}
up to linear order in the deformation parameter $\xi$.
The termination of the bulk polarization at the
surface of the cylinder induces a {\em bulk originated} surface
nonlinear charge,
$\sigma_{-,\text{src}}^b=\bm
P_{-,\text{src}}\cdot\hat{\bm n}$
where  $\hat{\bm n}=\hat{\bm
  r}+3\xi\sin3\theta\,\hat{\bm \theta}$ is the outwards pointing unit
vector normal to the surface. Substituting \cref{dfnwPnl} we obtain
\begin{align}
  \sigma_{-,\text{src}}^b=&\frac{E_1E_2^*\xi}
                                   {2\pi^2ne r_0}
               \frac{(1-\epsilon_1)(1-\epsilon_2^*)}
               {(1+\epsilon_1)^2(1+\epsilon_2^*)^2}
               \Bigg[-(1-\epsilon_{-})
               (2+\epsilon_1+\epsilon_2^*)
                +(1-\epsilon_1\epsilon_2^*)\nonumber\\
  &+\left(\frac{\omega_1+\omega_2}{\omega_{-}}\right)
    (\epsilon_2^*-\epsilon_1)
    \Bigg]\sin\theta.
    \label{dfnwsigmanlb}
\end{align}

We now turn our attention towards the surface of the cylinder and
calculate its nonlinear polarization. The expressions
for the normal and tangential component of the nonlinear surface DF
polarization are given by \cref{Psperp,Pspar}. Both of these
expressions were however written down for a semi-infinite surface lying
at $z=0$ with the $z$ direction towards the bulk. In
order to apply it to the curved
cylindrical surface, we assume the thickness of the selvedge to be
much smaller than the nominal radius $r_0$ of the
cylinder. This permits us to assume that the surface is locally flat
so the results of the dipolium model described in \cref{dfg}
become applicable. We also assume a local Cartesian system on the
surface with $\perp$ denoting the outwards pointing normal direction
and $\parallel$ denoting directions tangential to the
surface. The components of the nonlinear surface polarization
induced at each point on the surface can then be written as,
$P^s_{-,i}=\chi^s_{ijk}(\omega_1,\overline{\omega}_2)F_{1,j}F_{2,k}
+1\leftrightarrow 2$,
identical to \cref{chi-general}, where
${\chi}^{s}_{ijk}(\omega_1,\overline{\omega}_2)$ are the components
of the local nonlinear surface susceptibility and
\begin{equation}
  \bm F_g(r_s(\theta),\theta)=\bm E_g(r_s^+(\theta),\theta)=
  \epsilon_g\bm E_g^\perp(r_s^-(\theta),\theta)+
  \bm E_g^\parallel(r_s^-(\theta),\theta),
  \label{dfnwfieldf}
\end{equation}
where $r_s^\pm=r_s\pm\eta$, $\eta\to0^+$ are positions just outside
($+$) or within ($-$) the surface.
We recall that the fields $\bm F_g$ ($g=1,2$) are constant across
the thin selvedge.

Interpreting \cref{Psperp}
in the locally oriented frame we obtain the normal component of the surface
nonlinear polarization
\begin{align}\label{dfnwpsperp}
  P^s_{{_{-}},\perp}=&\frac{E_1E_2^*}{32\pi^2 ne}
                \frac{(1-\epsilon_1)(1-\epsilon_2^*)}
                {(1+\epsilon_1)(1+\epsilon_2^*)}\bigg[
  2a\sin2\theta + 6\xi a(\sin\theta+\sin5\theta)
  \nonumber\\
    +&8\xi a\sin3\theta
       \bigg\{\frac{\epsilon_1-\epsilon_2^*}
       {(1+\epsilon_1)(1+\epsilon_2^*)}\bigg\}
  +8\xi a\sin\theta
       \bigg\{\frac{1-\epsilon_1\epsilon_2^*}
       {(1+\epsilon_1)(1+\epsilon_2^*)}\bigg\}
  \bigg],
\end{align}
where the dimensionless parameters $a$ and $b$ are given by
\cref{aDF,b} respectively. Similarly, using \cref{Pspar} we obtain the
tangential component of the surface polarization. Its
spatial variation along the surface yields a {\em surface originated}
nonlinear surface charge
$\sigma_{-}^s=-\nabla_\|\cdot\bm{P}_{_{-},\parallel}^s$ given by
 \begin{align}
  \sigma_{-}^s=&\frac{E_1E_2^*}{16\pi^2 ne r_0}\,b\,
             \frac{(1-\epsilon_1)(1-\epsilon_2^*)}
             {(1+\epsilon_1)(1+\epsilon_2^*)}\bigg[
                    4\sin2\theta -4\xi\sin\theta +28\xi \sin5\theta
  \nonumber\\
  +&8\xi\frac{\epsilon_2^*-\epsilon_1}
     {(1+\epsilon_1)(1+\epsilon_2^*)}
     \bigg\{\frac{\omega_1+\omega_2}{\omega_{-}}
     \sin\theta-3\sin3\theta\bigg\}\bigg].\label{dfnwsigmas}
\end{align}

Now that we have calculated its sources, we turn our
attention to the calculation of the DF near field. The screened DF scalar
potential $\phi_{-}$ has
$\rho_{-,\text{src}} =0$ as an {\em external
  bulk} source and the total nonlinear charges induced at the surface
$\sigma_{-,\text{src}}^b$ and $\sigma_{-}^s$
as {\em surface} sources, together
with the normal component of the surface polarization
$P^s_{_{-},\perp}$,  which may be
accounted through the boundary conditions. The sources have to be screened by
the linear response $\epsilon_{-}$ of the particle at the DF
frequency. Thus, to obtain the quadratic self-consistent scalar
potential we have to solve
\begin{equation}\label{dfnwphi2}
\nabla^2\phi_{-}=
\begin{cases}
  0,& \text{(outside)}\\
  -4\pi \rho_{-,\text{src}}/\epsilon_{-}=0,& \text{(inside)}
\end{cases}
\end{equation}
subject to the boundary conditions
\begin{equation}
  \phi_{-}(r_s^+)-\phi_{-}(r_s^-)
  =4\pi P^s_{_{-},\perp},
  \label{dfnwbc21}
\end{equation}
and
\begin{equation}
  \hat{\bm n}\cdot\nabla\phi_{-}(r_s^+)
  -\epsilon_{-}
    \hat{\bm n}\cdot\nabla\phi_{-}(r_s^-)=-4\pi
      (\sigma_{-,\text{src}}^b+\sigma_{-}^s).
      \label{dfnwbc22}
\end{equation}
\cref{dfnwbc21} is the discontinuity of the scalar potential due to the
presence of the dipole layer
$P^{s}_{{_{-}},\perp}$
across the selvedge. \cref{dfnwbc22} is the discontinuity of the normal component of
the displacement field due to the presence of the nonlinear surface
charge. We solve \cref{dfnwphi2} perturbatively using
\cref{phitrial} (with the subscript $g=-$) to obtain the
self-consistent scalar potential at the DF frequency with terms up to
linear order in $\xi$.  The resulting DF self consistent
scalar potential $\phi^\mathrm{out}_-$ outside the cylinder is given by
\begin{align}\label{dfnwpoten}
  \frac{\phi^{\mathrm{out}}_{-}}{E_1E_2^*}=&
        \frac{\xi}{\pi ne}\frac{(1-\epsilon_1)(1-\epsilon_2^*)}
{(1+\epsilon_1)(1+\epsilon_2^*)(1+\epsilon_{-})}
  \bigg[\frac{2}{(1+\epsilon_1)(1+\epsilon_2^*)}\bigg\{
             -(1-\epsilon_{-})(2+\epsilon_1+\epsilon_2^*)
                                             \nonumber\\
  +&(1-\epsilon_1\epsilon_2^*)
+(\epsilon_2^*-\epsilon_1)\left(\frac{\omega_{1}+\omega_{2}}{\omega_{-}}\right)
                                           \bigg\}
     -b\frac{1+3\epsilon_{-}}{1+\epsilon_{-}}
    +2b\frac{\epsilon_2^*-\epsilon_1}{(1+\epsilon_1)(1+\epsilon_2^*)}
    \left(\frac{\omega_{1}+\omega_{2}}{\omega_{-}}\right)\nonumber\\
    +&\frac{\epsilon_{-}}{4}\frac{(3-\epsilon_{-})
   a}{1+\epsilon_{-}}
  +\frac{\epsilon_{-}}{4}\bigg\{3a+\frac{4a(1-\epsilon_1\epsilon_2^*)}
  {(1+\epsilon_1)(1+\epsilon_2^*)}\bigg\}\bigg]\frac{r_0}{r}\sin\theta
                                         &~\nonumber\\
                +&\frac{1}{4\pi ne}
                    \frac{(1-\epsilon_1)(1-\epsilon_2^*)}
{(1+\epsilon_1)(1+\epsilon_2^*)(1+\epsilon_{-})}
    [\epsilon_{-}a+2b]\frac{r_0^2}{r^2}\sin2\theta+\ldots
\end{align}
We must remark that terms corresponding to higher order multipoles are
present in this second order potential; however, they are of linear
order in deformation and smaller than the dipole by at least
$r_0/\lambda$. Comparing \cref{dfnwpoten} with the general expression
of the $2D$ scalar potential
\begin{equation}\label{phi2gendf}
  \phi^{\mathrm{out}}_-=2p_{-,y} \frac{\sin\theta}{r}
  +Q_{-,xy}\frac{\sin2\theta}{r^2}+\ldots
\end{equation}
we identify the DF 2D dipolar and quadrupolar moments per unit length
$p_{-,i}$ and $Q_{-,ij}$ The quadratic
dipole contributes only for a non-zero
deformation $\xi$ while the quadrupolar response $Q_{xy}$ is
independent of $\xi$ and would exist even for a non-deformed circular
cylinder.
We define dipolar and quadrupolar hyperpolarizability tensors through
$p_{-,i}=\gamma^d_{ijk}(\omega_1,\overline{\omega}_2)
E^{\mathrm{ex}}_{1,j}(E^{\mathrm{ex}}_{2,k})^*
+ 1\leftrightarrow2$
and
$Q_{-,ij}=\gamma^Q_{ijkl}(\omega_1,\overline{\omega}_2)
E^{\mathrm{ex}}_{1,k}(E^{\mathrm{ex}}_{2,l})^*
+ 1\leftrightarrow2$.
Together with \cref{phi2gendf}, they allow us to write the DF moments as
\begin{align}
  p_{-,y}=&[\gamma^d_{yxy}(\omega_1,\overline{\omega}_2)
        +\gamma^d_{yyx}(\overline{\omega}_2,\omega_1)]E_{1,x}E^*_{2,y}
        =\gamma^d(\omega_1,\overline{\omega}_2)
        E_{1,x}E^*_{2,y},\label{DFnld}\\
  Q_{-,xy}=&[\gamma^Q_{xyxy}(\omega_1,\overline{\omega}_2)
           +\gamma^Q_{xyyx}(\overline{\omega}_2,\omega_1)]E_{1,x}E^*_{2,y}
           =\gamma^Q(\omega_1,\overline{\omega}_2)
           E_{1,x}E^*_{2,y},\label{DFnlQ}
\end{align}
where we define
\begin{equation}
\gamma^d(\omega_1,\overline{\omega}_2)=
  \gamma^d_{yxy}(\omega_1,\overline{\omega}_2)
+\gamma^d_{yyx}(\overline{\omega}_2,\omega_1)
=2\gamma^d_{yxy}(\omega_1,\overline{\omega}_2)
=2\gamma^d_{yyx}(\overline{\omega}_2,\omega_1)\label{DFnlgammad}
\end{equation}
and
\begin{equation}
\gamma^Q(\omega_1,\overline{\omega}_2)=
  \gamma^Q_{xyxy}(\omega_1,\overline{\omega}_2)
+\gamma^Q_{xyyx}(\overline{\omega}_2,\omega_1)
=2\gamma^Q_{xyxy}(\omega_1,\overline{\omega}_2)
=2\gamma^Q_{xyyx}(\overline{\omega}_2,\omega_1)\label{Dfnlgammaq}
\end{equation}
using the intrinsic permutation symmetry
$\gamma^d_{yxy}(\omega_1,\overline{\omega}_2)=
\gamma^d_{yyx}(\overline{\omega}_2,\omega_1)$ and
$\gamma^Q_{xyxy}(\omega_1,\overline{\omega}_2)
=\gamma^Q_{xyyx}(\overline{\omega}_2,\omega_1)$. Comparing
\cref{DFnld,DFnlQ} with
\cref{phi2gendf,dfnwpoten}, we identify the DF hyperpolarizabilities
\begin{align}\label{dfgammad}
  \gamma^d(\omega_1,\overline{\omega}_2)&=\frac{\xi r_0}{2\pi ne}
                                 \frac{(1-\epsilon_1)(1-\epsilon_2^*)}
{(1+\epsilon_1)(1+\epsilon_2^*)(1+\epsilon_{-})}
  \bigg[\frac{2}{(1+\epsilon_1)(1+\epsilon_2^*)}\bigg\{
             -(1-\epsilon_{-})(2+\epsilon_1+\epsilon_2^*)\nonumber\\
  +&(1-\epsilon_1\epsilon_2^*)
+(\epsilon_2^*-\epsilon_1)\left(\frac{\omega_{1}+\omega_{2}}{\omega_{-}}\right)
                                           \bigg\}
     - b\frac{1+3\epsilon_{-}}{1+\epsilon_{-}}
    +2 b\frac{(\epsilon_2^*-\epsilon_1)}{(1+\epsilon_1)(1+\epsilon_2^*)}
    \left(\frac{\omega_{1}+\omega_{2}}{\omega_{-}}\right)\nonumber\\
    +&\frac{\epsilon_{-}}{4}\frac{(3-\epsilon_{-})
   a}{1+\epsilon_{-}}
  +\frac{\epsilon_{-}}{4}\bigg\{3a+\frac{4a(1-\epsilon_1\epsilon_2^*)}
     {(1+\epsilon_1)(1+\epsilon_2^*)}\bigg\}\bigg],
\end{align}
\begin{align}\label{dfgammaq}
  \gamma^Q(\omega_1,\overline{\omega}_2)= \frac{r_0^2}{4\pi ne}
                    \frac{(1-\epsilon_1)(1-\epsilon_2^*)}
{(1+\epsilon_1)(1+\epsilon_2^*)(1+\epsilon_{-})}
    (a\epsilon_{-}+2b).
\end{align}

We now turn our attention towards the calculation of the nonlinear
magnetic dipole moment induced in the nanocylinder, the density of
which is given by \cref{DFmag}. The total magnetic moment per unit
length induced within the nanocylinder can be obtained by integrating
\cref{DFmag} across the cross-section,
\begin{equation}\label{nlmagdip}
\bm{m}_{-}^{(2)}=\int^{2\pi}_{0}\int^{{r}_s(\theta)}_0
\bm{\mu}_{-}^{(2)} r dr d\theta.
\end{equation}
Substituting the linear
fields obtained from \cref{dfnwphi12} in
\cref{DFmag}, we
obtain
\begin{align}\label{DFmagmom}
  \bm{m}_{-}^{(2)}=-\frac{i}{c e}\frac{E_1E_2^*\,r_0^2}{8\pi n^2}
  (\omega_1+\omega_2)
  \frac{(1-\epsilon_1)(1-\epsilon_2^*)}{(1+\epsilon_1)(1+\epsilon_2^*)}
  \bm{\hat{z}}.
\end{align}
Defining
$m_{-,z}=\gamma^m_{zxy}(\omega_1,\overline{\omega}_2)E_{1,x}E_{2,y}^*
+1\leftrightarrow2$, and following a procedure similar to
\cref{DFnld}, we identify the quadratic magnetic dipolar
hyperpolarizability
$\gamma^m(\omega_1,\overline{\omega}_2)$. We obtain
\begin{align}\label{DFgammam}
  \gamma^m(\omega_1,\overline{\omega}_2)
  =2\gamma^m_{zxy}(\omega_1,\overline{\omega}_2)
  =-\frac{i}{c e}\frac{r_0^2}{8\pi n^2}
  (\omega_1+\omega_2)
  \frac{(1-\epsilon_1)(1-\epsilon_2^*)}{(1+\epsilon_1)(1+\epsilon_2^*)}.
\end{align}

The limit $\omega_2\to\omega_1$, $\omega_{-}\to0$ of
\cref{dfgammad,dfgammaq,DFgammam} yields
the OR hyperpolarizabilities, i.e. the
second order response of the nanocylinder for the nonlinear
rectification process. The SF hyperpolarizabilities, corresponding to
the frequency $\omega_{+}=\omega_1+\omega_2$, can be easily read
from \cref{dfgammad,dfgammaq,DFgammam} after substituting
$\overline{\omega}_2$ by $\omega_2$, $\epsilon_{-}$ by $\epsilon_{+}$,
and taking the complex conjugate of the permittivity, i.e. changing
$\epsilon_2^*$ to $\epsilon_2$, yielding
\begin{align}\label{sfgammad}
  \gamma^d(\omega_1,\omega_2)&=\frac{\xi r_0}{2\pi ne}
                                 \frac{(1-\epsilon_1)(1-\epsilon_2)}
{(1+\epsilon_1)(1+\epsilon_2)(1+\epsilon_{+})}
  \bigg[\frac{2}{(1+\epsilon_1)(1+\epsilon_2)}\bigg\{
             -(1-\epsilon_{+})(2+\epsilon_1+\epsilon_2)\nonumber\\
                              +&(1-\epsilon_1\epsilon_2)
+(\epsilon_2-\epsilon_1)\left(\frac{\omega_{1}-\omega_{2}}{\omega_{+}}\right)
                                           \bigg\}
     -b\frac{1+3\epsilon_{+}}{1+\epsilon_{+}}
    +2 b\frac{\epsilon_2-\epsilon_1}{(1+\epsilon_1)(1+\epsilon_2)}
    \left(\frac{\omega_{1}-\omega_{2}}{\omega_{+}}\right)\nonumber\\
    +&\frac{\epsilon_{+}}{4}\frac{(3-\epsilon_{+})
   a}{1+\epsilon_{+}}
  +\frac{\epsilon_{+}}{4}\bigg\{3a+\frac{4a(1-\epsilon_1\epsilon_2)}
     {(1+\epsilon_1)(1+\epsilon_2)}\bigg\}\bigg],
\end{align}
\begin{align}\label{sfgammaq}
  \gamma^Q(\omega_1,\omega_2)= \frac{r_0^2}{4\pi ne}
                    \frac{(1-\epsilon_1)(1-\epsilon_2)}
{(1+\epsilon_1)(1+\epsilon_2)(1+\epsilon_{+})}
    [\epsilon_{+}a+2b],
\end{align}
and
\begin{align}\label{sfgammam}
  \gamma^m(\omega_1,{\omega}_2)
  =-\frac{i}{c e}\frac{r_0^2}{8\pi n^2}
  (\omega_1-\omega_2)
  \frac{(1-\epsilon_1)(1-\epsilon_2)}{(1+\epsilon_1)(1+\epsilon_2)}.
\end{align}

The degenerate SH case can be obtained from \cref{sfgammad,sfgammaq}
when the input frequencies are equal, i.e. $\omega_2=\omega_1$. Note,
that the magnetic hyperpolarizability given by \cref{sfgammam} would
be zero for the SH case.

In order to calculate the other non-zero components of the
hyperpolarizabilities, we repeat the
calculations above but with different polarization of the incident
fields to find all the non-zero
components of the hyperpolarizabilities,
%
%
\begin{equation}\label{dfgamdsymv2}
  \begin{split}
    \gamma^d_{bab}(\omega_c,\omega_d)&
    =\gamma^d_{bba}(\omega_c,\omega_d)
    =-\gamma^d_{aaa}(\omega_c,\omega_d)
    =\gamma^d_{abb}(\omega_c,\omega_d)\\
    &=\gamma^d(\omega_c,{\omega}_d)/2
    =\gamma^d({\omega}_d,\omega_c)/2,
  \end{split}
\end{equation}
%
%
%
\begin{align}\label{dfgamQsym}
  \begin{split}
  \gamma^Q_{abab}(\omega_c,\omega_d)
  &=\gamma^Q_{abba}(\omega_c,\omega_d)
  =-\gamma^Q_{aabb}(\omega_c,\omega_d)
  =\gamma^Q_{aaaa}(\omega_c,\omega_d)\\
  &=\gamma^Q(\omega_c,\omega_d)/2 =\gamma^Q(\omega_d,\omega_c)/2,
  \end{split}
\end{align}
%
%
%
%
\begin{equation}
\gamma^m_{zab}(\omega_c,{\omega}_d)
=\gamma^m_{zba}({\omega}_d,\omega_c)
=\gamma^m(\omega_c,{\omega}_d)/2
=-\gamma^m(\omega_d,{\omega}_c)/2
\label{dfgammsymv2}
\end{equation}
where the pair of indices
$(a,b)$ can take the values $(x,y)$ or $(y,x)$, and the pair of
frequencies $(\omega_c,{\omega}_d)$ can take independently the values
$(\omega_1,\omega_2)$ or $(\omega_2,{\omega}_1)$.
All other components are zero for our system.

\subsection{SF/DF efficiency}\label{dfconveff}

We now focus on the calculation of the electromagnetic fields in
the radiation zone and the efficiency of DFG/SFG from the
nanocylinder. Following a procedure similar to the $3D$ case, one
can write down the expressions for the radiated electromagnetic fields
in $2D$ due to a localized distribution of charges and
currents \cite{raksha}. The
magnetic and electric far fields radiated at the SF/DF frequency are
\begin{equation}
  \bm B_{\pm}=(1+i)k_{\pm}^{3/2}\left((\hat{\bm r}\times \bm p_{\pm})
    -\bm{\hat{r}}\times(\bm{\hat{r}}\times\bm{m}_{\pm})
    -\frac{i}{4} k_{\pm} (\hat{\bm r}
    \times(\bm Q_{\pm}\cdot\hat{\bm r}))\right)
  e^{ik_{\pm}
    r}\sqrt{\frac{\pi}{r}},
  \label{Brad}
\end{equation}
\begin{equation}
  \bm E_{\pm}=\bm B_{\pm}\times\hat{\bm r},
  \label{Erad}
\end{equation}
considering the dominant electric-dipolar, magnetic-dipolar, and
electric-quadrupolar contributions,  where $k_{\pm}$ is the free wavenumber
corresponding to the frequency
$\omega_\pm$, $\hat{\bm r}$ is the outward pointing unit vector
in the direction of observation, and
$\bm{p}_{\pm}$, $\bm{Q}_{\pm}$ and  $\bm{m}_{\pm}$ are given by
\cref{DFnld,DFmagmom,DFnlQ} respectively. In the appendix of
Ref. \cite{raksha} the electric dipolar and quadrupolar contributions
to the radiated fields were presented; the magnetic dipolar
contribution is discussed in our Appendix. The time averaged
power radiated per unit angle in the direction $\theta$ is
\begin{equation}
  \frac{dP_{\pm}}{d\theta}=\frac{rc}{2\pi}
  \operatorname{Re}[\bm E_{\pm} \times \bm B_{\pm}^{*}]\cdot \hat{\bm r},
  \label{angrad}
\end{equation}
which after substituting \cref{Brad,Erad} becomes
\begin{align}\label{dfangrad}
  \frac{d\mathcal{P}_{\pm}}{d\theta}=c k_{\pm}^3|E_1E_2|^2
  \big[|\gamma^d_{\pm}|^2&\cos^2\theta+|\gamma^m_{\pm}|^2
    +\frac{k_{\pm}^2}{16}|\gamma^Q_{\pm}|^2\cos^22\theta
    +2\operatorname{Re}(\gamma^d_{\pm}\gamma^{m*}_{\pm})\cos\theta
    \nonumber\\
        &-\frac{k_{\pm}}{2}\operatorname{Im}(\gamma^m_{\pm}\gamma^{Q*}_{\pm})
  \cos2\theta
    -\frac{k_{\pm}}{2}\operatorname{Im}(\gamma^d_{\pm}\gamma^{Q*}_{\pm})
  \cos\theta\cos2\theta\big].
\end{align}
Here, we introduce the compact notation
$\gamma^\alpha_+\equiv\gamma^\alpha(\omega_1,\omega_2)$ and
$\gamma^\alpha_-\equiv\gamma^\alpha(\omega_1,\overline \omega_2)$ with
$\alpha=d,m,Q$. Integrating we
obtain the total SF/DF power radiated per unit length,
\begin{equation}\label{dfpower}
  \mathcal{P}_{\pm}=\pi c
  k_{\pm}^3|E_1E_2|^2\left[|\gamma^d_{\pm}|^2
    +2|\gamma^m_{\pm}|^2+
  \frac{k_{\pm}^2}{16}|\gamma^Q_{\pm}|^2\right].
\end{equation}
The efficiency of the SFG/DFG process, defined as
\begin{equation}\label{effdef}
  \mathcal{R}_{\pm}=\frac{\mathcal{P}_{\pm}}{I_1I_2},
\end{equation}
where $I_g=(c/2\pi)|E_g|^2$ ($g=1,2$) are the intensities of the
incident waves, is
\begin{equation}\label{dfgeff}
  \mathcal{R}_{\pm}=\frac{\pi^3}{1024}\frac{k_{\pm}^3}{c}\left[
    |\gamma^d_{\pm}|^2+2|\gamma^m_{\pm}|^2
    +\frac{k_{\pm}^2}{16}|\gamma^Q_{\pm}|^2\right],
\end{equation}
%

\section{Results and discussions}\label{results}
\begin{figure}
  \hspace{-0.7cm}
  \begin{tabular}{c c}
  \includegraphics[width=0.6\linewidth]{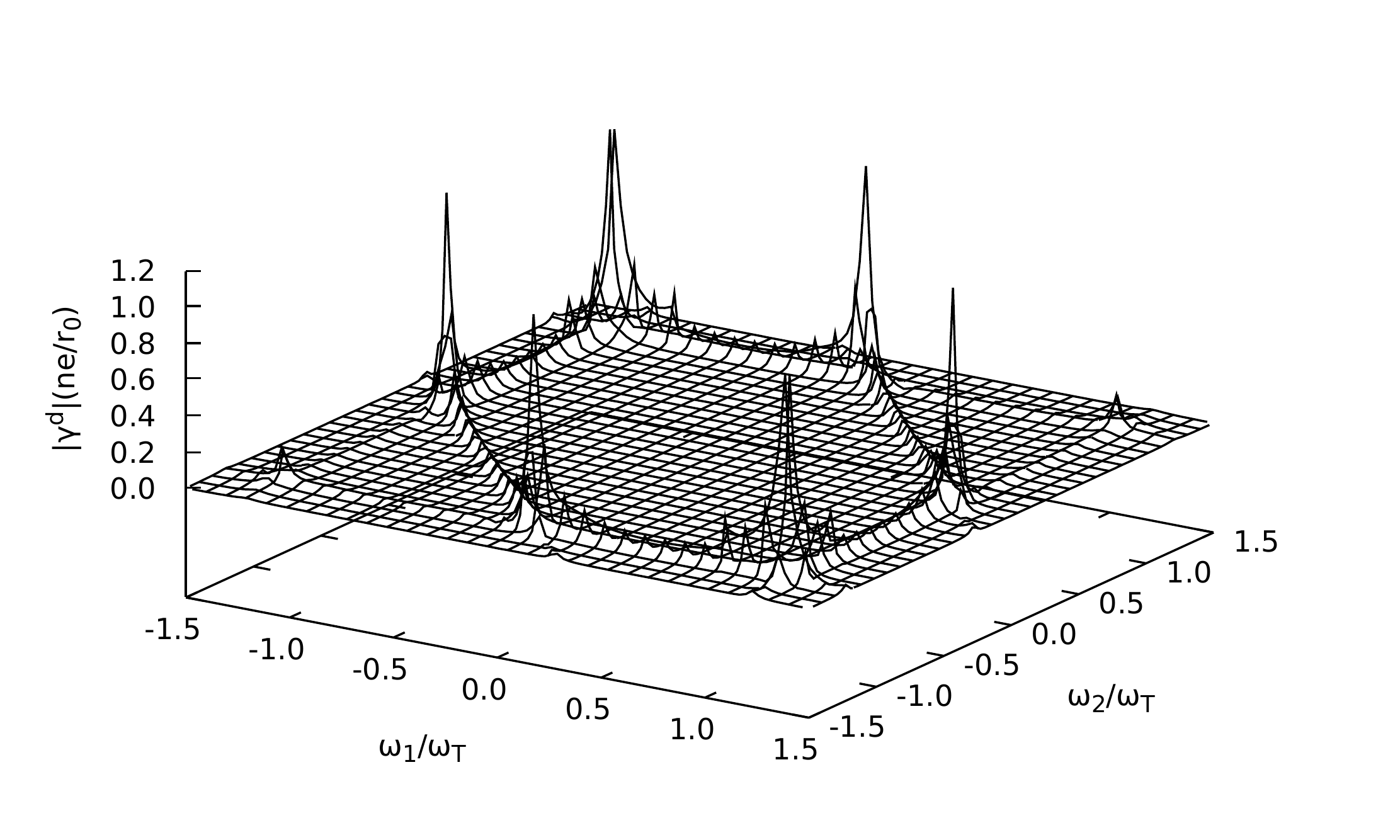}&
  \hspace{-1.2cm}
  \includegraphics[width=0.3\linewidth]{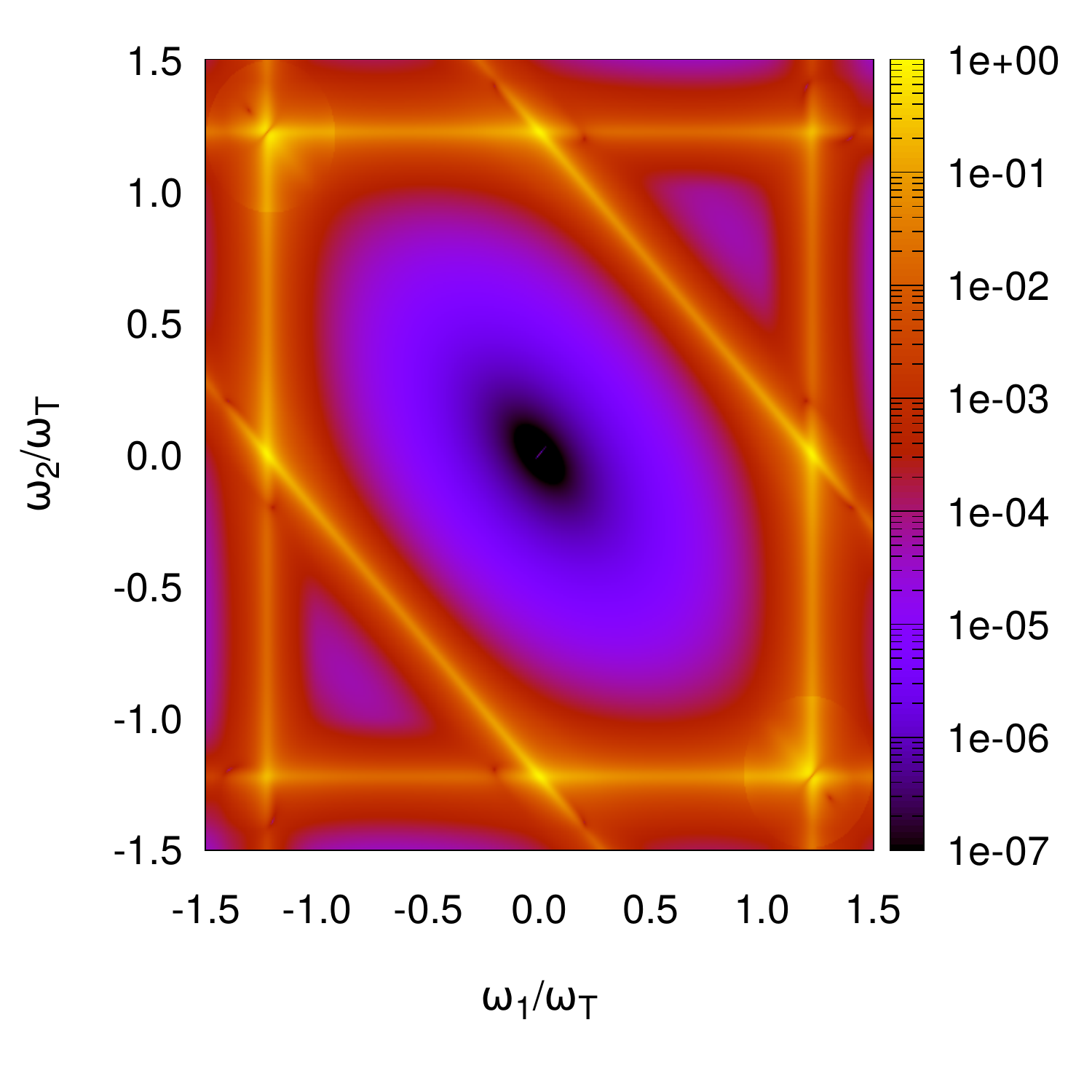}\\
  \includegraphics[width=0.6\linewidth]{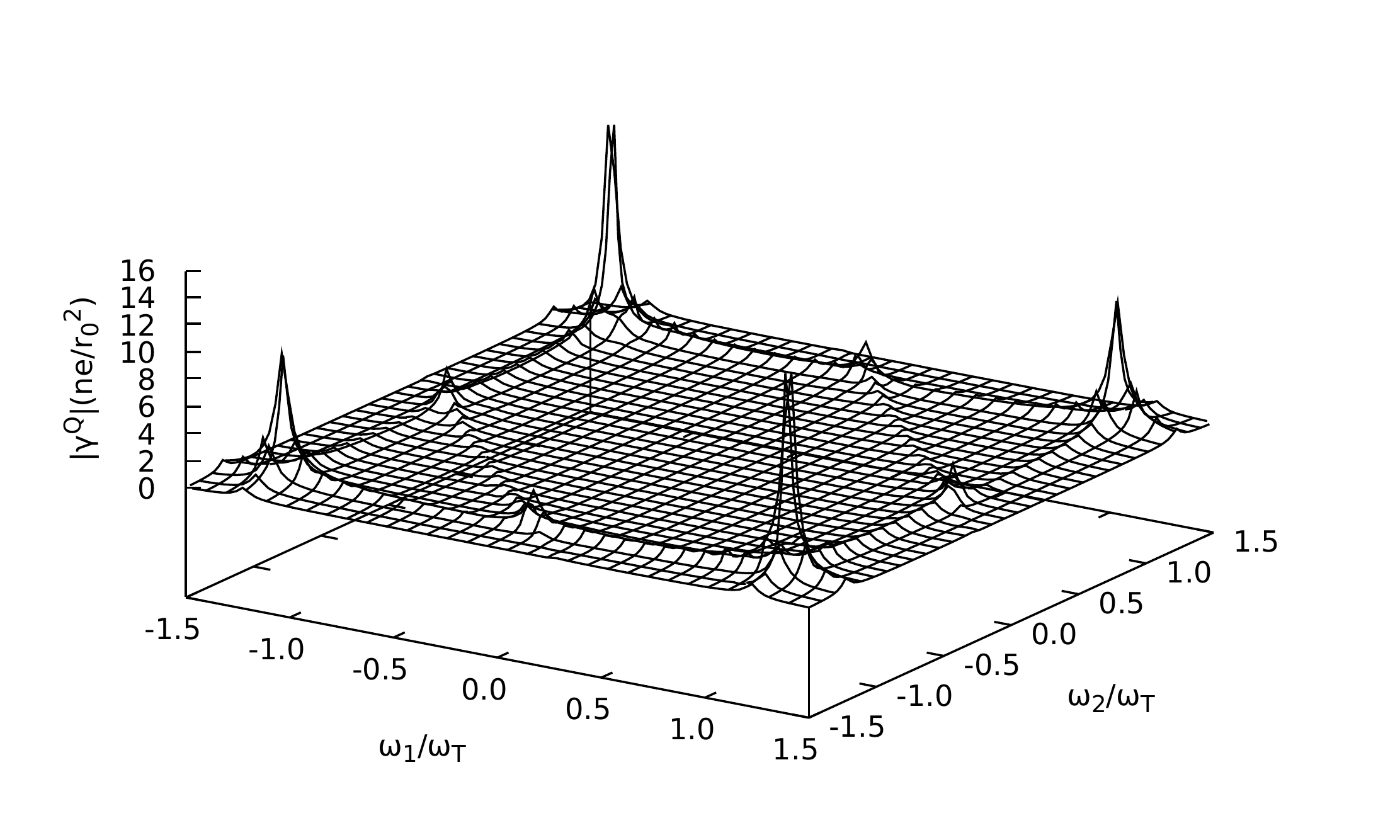}&
  \hspace{-1.2cm}
  \includegraphics[width=0.3\linewidth]{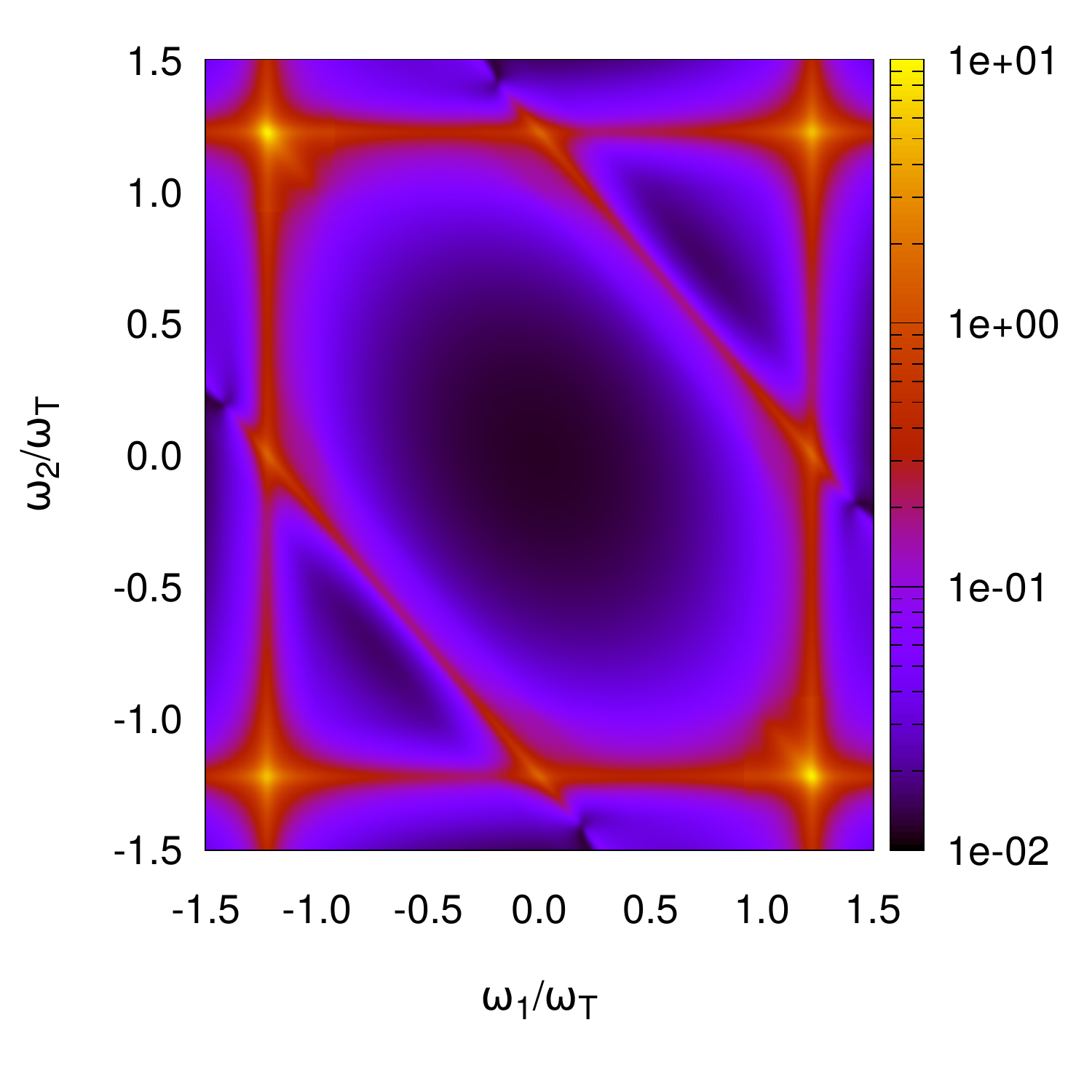}\\
  \includegraphics[width=0.6\linewidth]{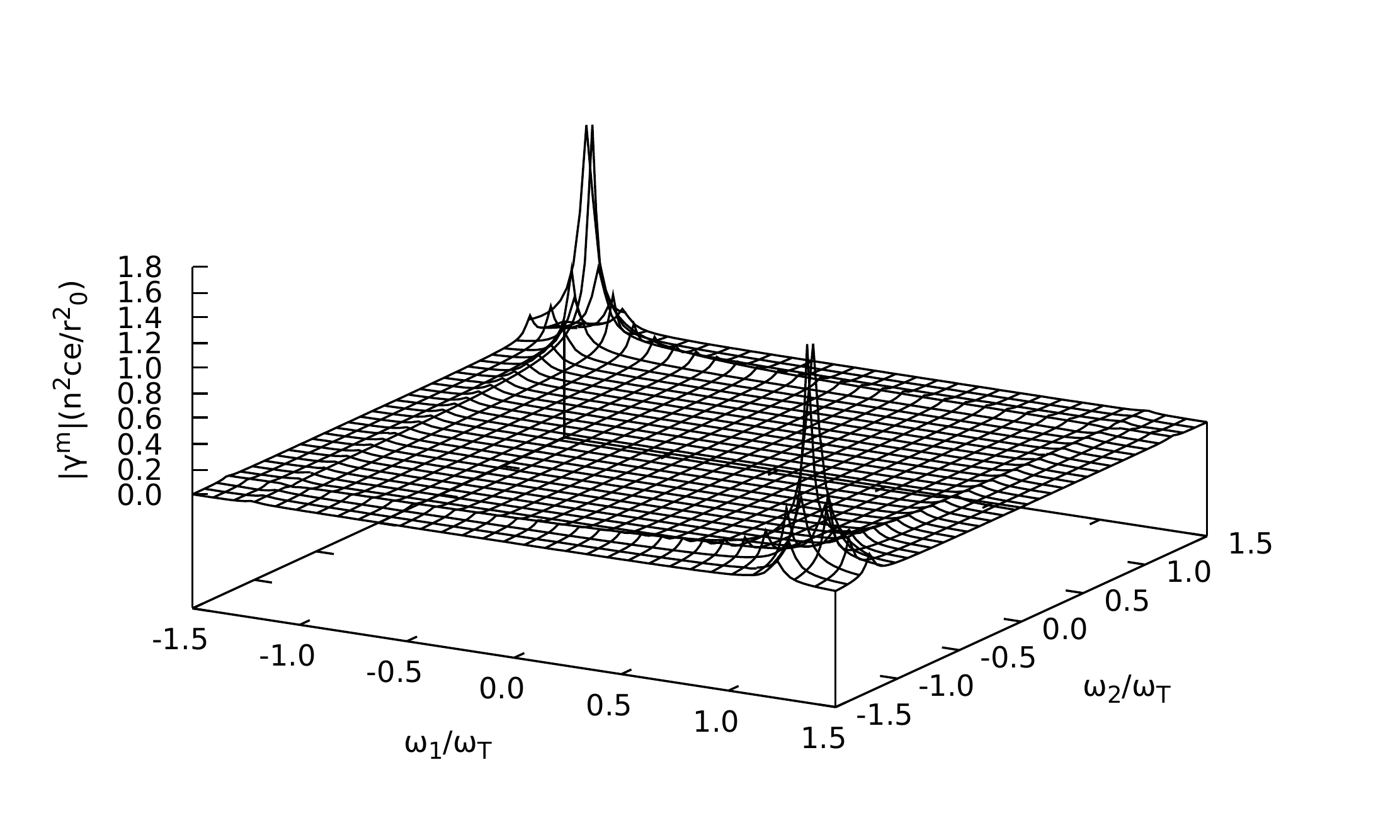}&
  \hspace{-1.2cm}
  \includegraphics[width=0.3\linewidth]{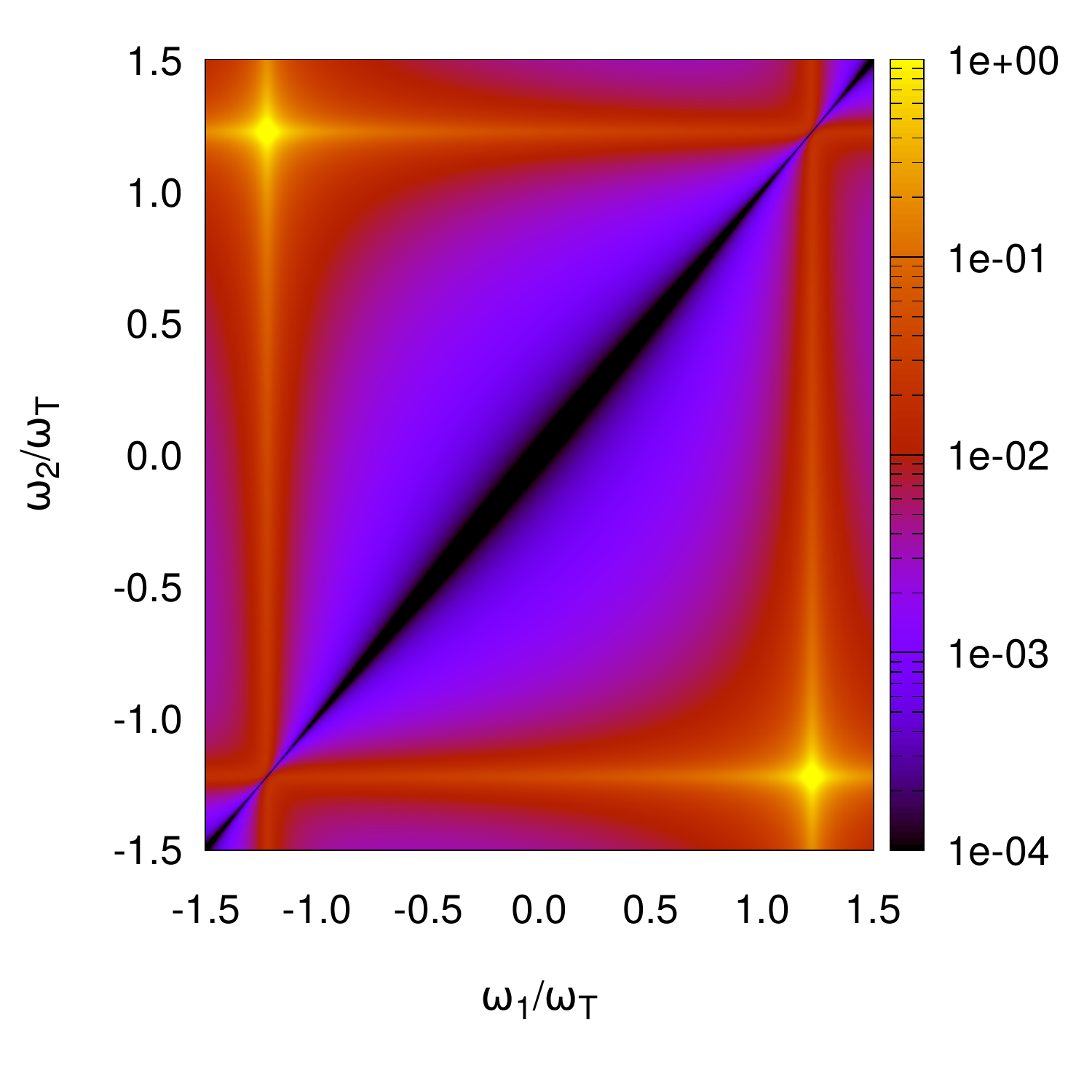}
  \end{tabular}
  \caption{\label{dfdielgamdQm}
     Normalized absolute value of the electric dipolar $\gamma^d$ (top
     panels), the electric quadrupolar (middle panels) $\gamma^Q$ and
     the magnetic dipolar (bottom panels) $\gamma^m$
     hyperpolarizabilities for an infinitely long and thin deformed dielectric
     cylinder as a function of the fundamental frequencies $\omega_1$
     and $\omega_2$. The permittivity is given by \cref{gendiel} with
     $\omega_{\mathrm{L}}=\sqrt{2}\,\omega_{\mathrm{T}}$
     and $\tau=20/\omega_{\mathrm{T}}$. The deformation is
     $\xi=0.03$. $3D$ surface plots are displayed in the left panels
     and the respective $2D$ color maps
     in the right panels. The regions where both
     frequencies have the same signs correspond to SFG and those with
     opposite signs to DFG.}
\end{figure}
\begin{figure}
  \center
  \includegraphics[width=0.6\linewidth]{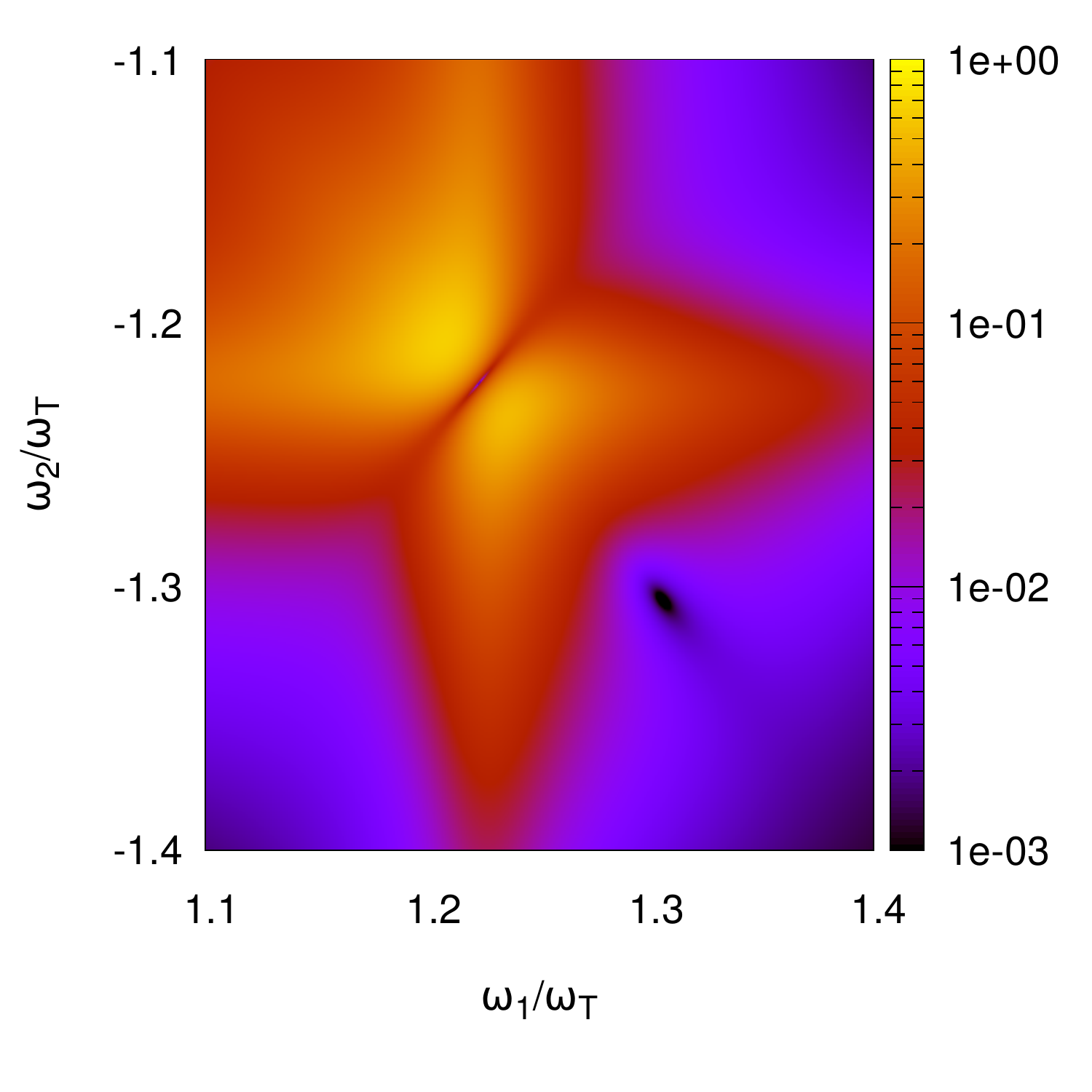}
  \caption{\label{dfgamd-peak}
    High resolution $2D$ color map of the normalized dipolar
    hyperpolarizability $|\gamma^d|$, shown in
    \cref{dfdielgamdQm},  close to the resonance around the region
    $\omega_1\approx\-\omega_2\approx\omega_{\mathrm{spp}}$, with
    $\omega_2$ negative.}
\end{figure}

\cref{dfdielgamdQm} illustrates the absolute value of the
electric-dipolar, electric-quadrupolar, and the
magnetic-dipolar hyperpolarizabilities $\gamma^\alpha$, $\alpha=d,Q,m$
given by \cref{dfgammad,dfgammaq,DFgammam,sfgammad,sfgammaq,sfgammam}
respectively, for a deformed cylinder made up of an insulator with
dielectric permittivity given by \cref{gendiel} with
$\omega_{\mathrm{L}}=\sqrt2\,\omega_{\mathrm{T}}$  and
$\tau=20/\omega_{\mathrm{T}}$, as used in \cref{reim-a-df}. We allow
both the input frequencies to take negative and positive values to
cover all three-wave mixing processes, namely, SFG, DFG, SHG, and OR,
by identifying
$\gamma^\alpha(\omega_1,\omega_2)=\gamma^\alpha_{+}$ when
both frequencies are positive,
$\gamma^\alpha(\omega_1,\omega_2)=\gamma^\alpha_{-}$ when
$\omega_1>0$ and $\omega_2<0$, and
$\gamma^\alpha(\omega_1,\omega_2)=(\gamma^\alpha(-\omega_1,-\omega_2))^*$ when
$\omega_1<0$.
We show both $3D$ surface plots and $2D$ color maps to better convey
the qualitative and quantitative nature of the results. All three
hyperpolarizabilities show strong resonant ridges when
either input frequency is equal to the
surface plasmon polariton (SPP) frequency or its
additive inverse,
$\omega_g=\pm \omega_{\mathrm{spp}}=\pm \sqrt{3/2}\,\omega_{\mathrm{T}}$. Intense
diagonal ridges occur for
$\gamma^d$ and $\gamma^Q$, but not in $\gamma^m$, when the sum of the two incident
frequencies resonates with the SPP. There are
further peaks when any of the two ridges meet, for which two of the
resonant conditions are fulfilled jointly. The first quadrant
corresponds to SFG, the fourth with DFG, and
the third and second replicate these
processes inverting the signs of all participating frequencies. The
large peak observed in the fourth quadrant corresponds
to DFG close to OR,  where both the incident frequencies are
simultaneously SPP resonant. The response along the diagonal
$\omega_1=\omega_2$ corresponds to SHG. Notice that in this
case we cross a diagonal ridge when the fundamental frequency is the
subharmonic of the SPP resonance and meet a peak when the fundamental
reaches the resonance condition. The quadratic magnetic dipole is
absent along the SH line. The horizontal,
vertical, and diagonal ridges are much weaker than the doubly
resonant peaks for both the quadrupolar and the dipolar
response. We also explored the absolute values of $\gamma^d$,
$\gamma^Q$ and $\gamma^m$ for larger lifetimes (not
shown). As expected, we obtained a similar structure with much
narrower and sharper peaks and ridges. In
\cref{dfgamd-peak}, we show a closeup of $\gamma^d$
around $\omega_1=-\omega_2=\omega_{\mathrm{spp}}$,
corresponding to DFG process with a small difference in frequency. Notice
that when the two input frequencies
are exactly equal and opposite, $\omega_-=0$, the second
order response is about an order of magnitude lower than for
neighboring points, for which $\omega_-$ is small but finite. Thus, the
electric-dipolar response of the system for OR is smaller than its DF
response. No such
behavior is observed for the quadrupolar nor the magnetic
dipolar hyperpolarizability.

\begin{figure}
  \center
  \includegraphics[width=0.65\linewidth]{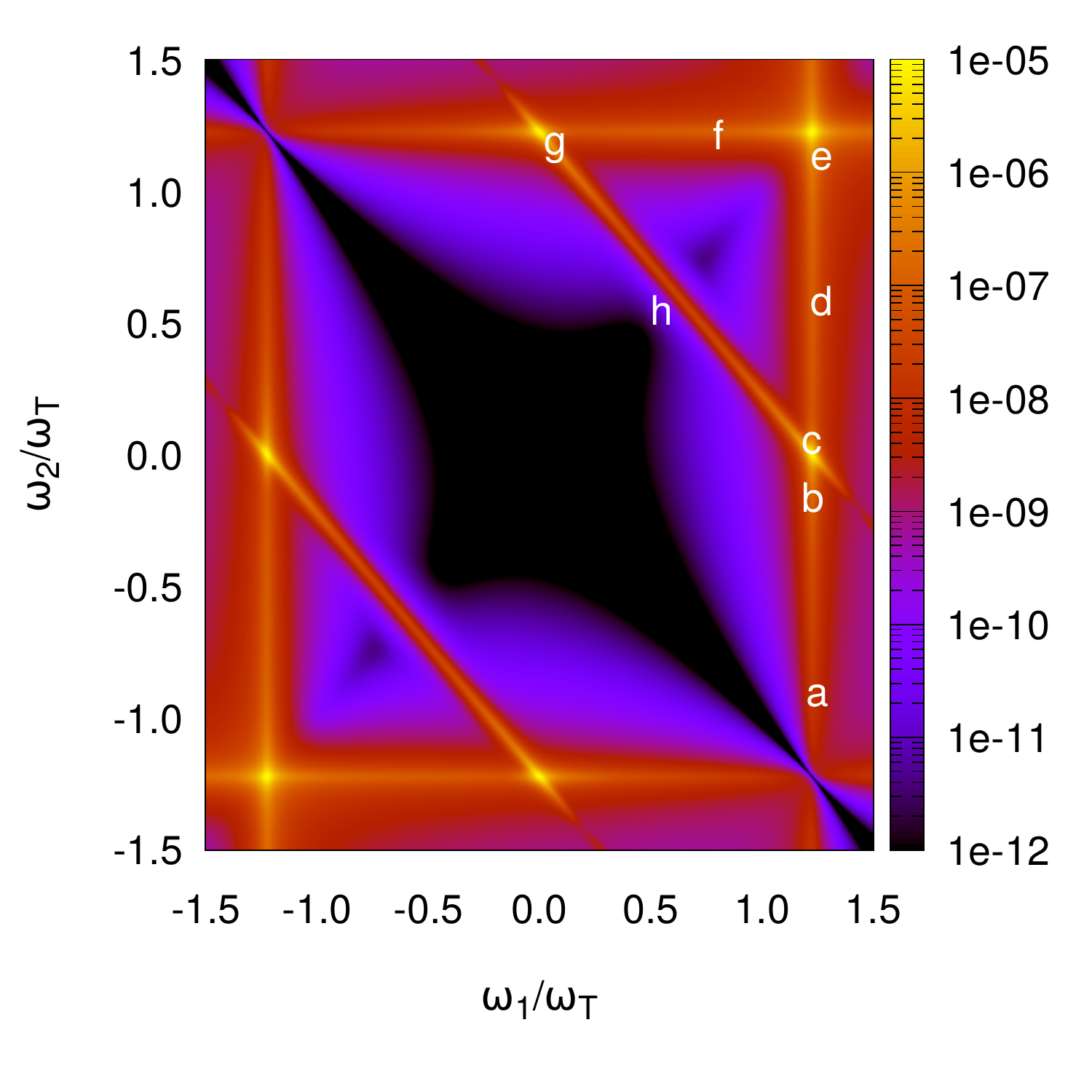}\\
  \caption{\label{dielsfdfeff}
     Dimensionless efficiency for three-wave mixing processes for the same
     dielectric cylinder as in \cref{dfdielgamdQm} as function
     of normalized input frequencies $\omega_g/\omega_{\mathrm{T}}$
     for $g=1,2$ for deformation $\xi=0.03$ and
     $r_0/\lambda_\mathrm{T}=0.01$.}
\end{figure}

In \cref{dielsfdfeff} we
present the SFG/DFG dimensionless efficiencies
\begin{equation}\label{dfnormeff}
  \mathcal{R}'_{\pm}=cr_0 (ne)^2 \mathcal{R}_{\pm}
\end{equation}
corresponding to the same
cylinder as in \cref{dfdielgamdQm}. We obtain a
structure similar to that for the hyperpolarizabilities, with vertical,
horizontal and diagonal ridges and peaks where two ridges meet. Note
that the main peak correspond to a SHG process where both input
frequencies are resonant with the SPP of the cylinder, followed in
intensity by SFG/DFG peaks where one input frequency is close to zero. The
SHG at the SPP subharmonic is relatively small and there is no peak
corresponding to OR where both frequencies are SPP
resonant. Nevertheless, there is a substantial DFG radiation when one
frequency is SPP resonant and the other is close to the resonance,
indicating a possible application towards the generation of THz
radiation.

\begin{figure}
  \center
  \includegraphics[width=0.65\linewidth]{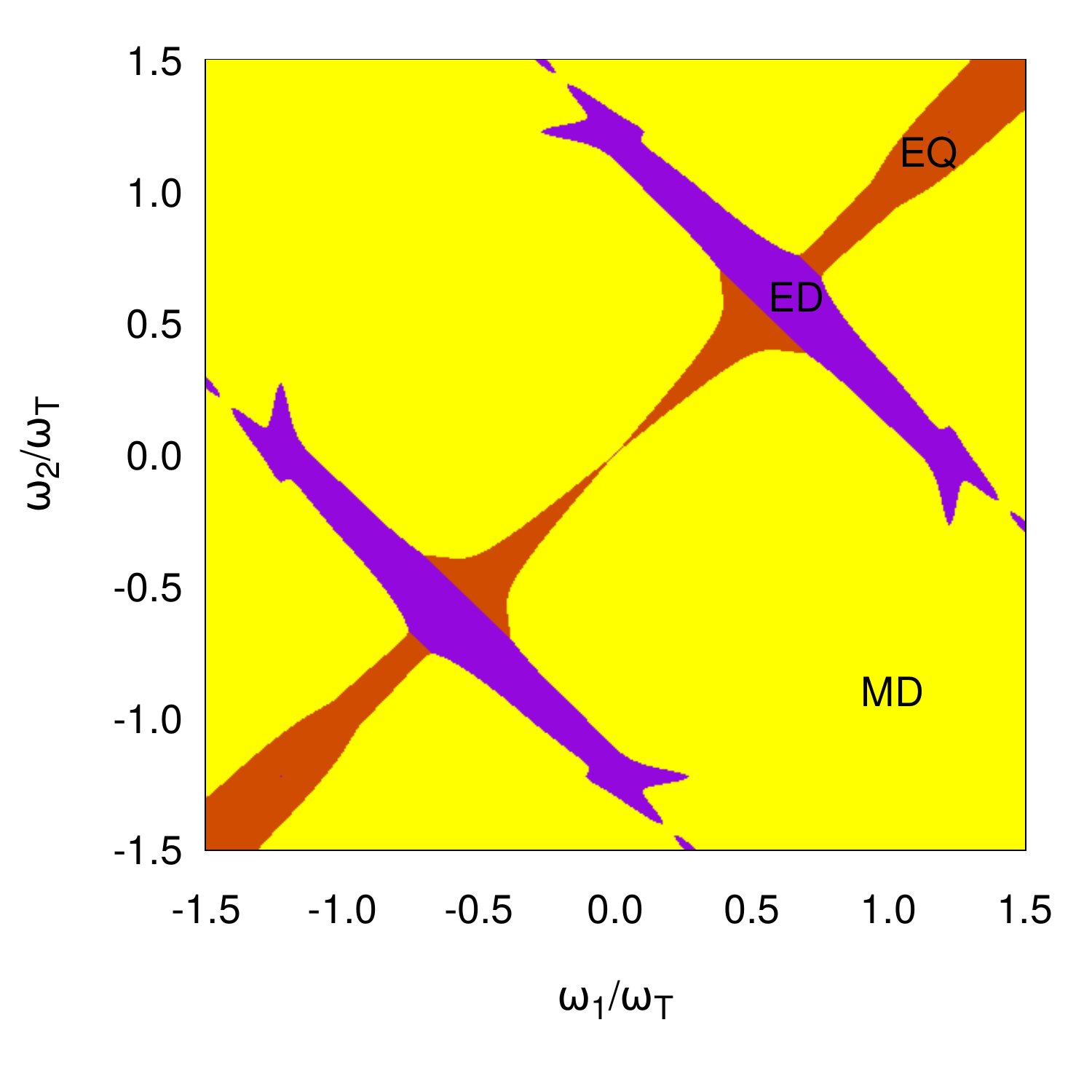}\\
  \caption{\label{contrib}
    Different regions of frequency space colored according to the
    nature of the largest contribution to the non-linear efficiency shown in
    \cref{dielsfdfeff}, electric-dipolar (ED), magnetic-dipolar (MD),
    and electric-quadrupolar (EQ).}
\end{figure}

In \cref{contrib} we show the
regions of frequency ($\omega_1,\omega_2$) space where the largest
contribution to the
nonlinear efficiency is electric-dipolar, magnetic-dipole or
electric quadrupolar. Notice that the electric dipole is dominant when
$\omega_+$ resonates with the SPP, and that the electric quadrupole
dominates along the SHG line, except at the subharmonic of the
SPP. The rest of the frequency space is dominated by the magnetic
dipole.

\begin{figure}
  \center
  \includegraphics[width=0.8\linewidth,trim={0 0 0 3cm},clip]
  {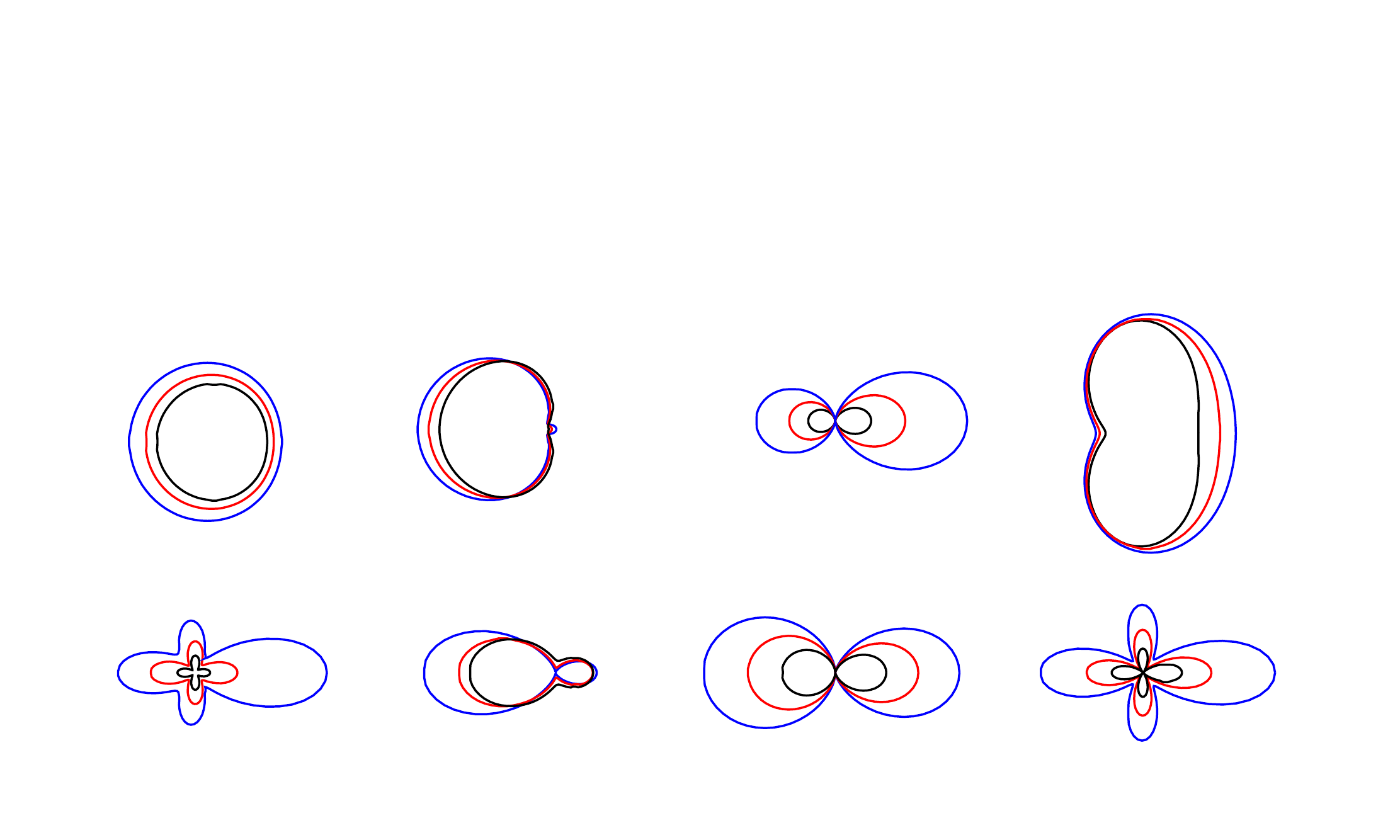}
  \caption{\label{radpatdiel}
     $2D$ angular radiation patterns for the same dielectric
     cylinder as in \cref{dielsfdfeff} for different input frequencies
     $\omega_1$ and $\omega_2$. Each set of patterns corresponds to the
   region around the points marked in \cref{dielsfdfeff}: from left to
   right a-d (top row), and e-h (bottom row). For the patterns
 a-e we set $\omega_1=\omega_{\mathrm{spp}}$ while
 varying $\omega_2$, i.e. we traverse along the vertical line at
 resonance, f lies on the horizontal line where
 $\omega_2=\omega_{\mathrm{spp}}$ . We choose both the input
 frequencies to vary along the diagonal resonance for regions g and h.}
\end{figure}

In \cref{radpatdiel}, we plot the
normalized $2D$ angular radiation pattern of the same cylinder as in
\cref{dfdielgamdQm} and \cref{dielsfdfeff} in the vicinity of resonances
corresponding to the regions marked in \cref{dielsfdfeff}. Even though
the calculation corresponds to a
small deformation, $\xi=0.03$, we observe a strong competition between
the electric-dipolar, magnetic-dipolar, and quadrupolar
contributions. Thus, around the region $a$ of Fig. \ref{dielsfdfeff},
in Fig. \ref{radpatdiel} we obtain an almost isotropic radiation
pattern along the $x-y$ plane, corresponding to the radiation of a
magnetic dipole oriented along $z$. As we move vertically towards
point $b$ there is a competition between electric and magnetic dipolar
radiation, which yields a partially directional radiation and as we
approach $c$ the radiation has mostly the two lobed
electric-dipolar form. As we proceed to $d$ the radiation becomes a
mixture
of magnetic-dipolar and electric-quadrupolar and the latter dominates
close to $e$. Moving towards $f$ and $g$ the pattern becomes again
electric-dipolar and finally, close to $h$ it becomes a mixture of
electric dipolar and quadrupolar contributions. We must remark here
that though \cref{dielsfdfeff} is symmetric under the interchange
$\omega_1\leftrightarrow\omega_2$, we purposely chose
points $c$ and $g$ that are not equivalent, in order to show the
richness of the radiation patterns. Similarly, we chose $d$
nonequivalent to $f$. An animation illustrating the
evolution of the generated radiation pattern as we vary continuously the  input
frequencies is available in the supplementary video file\cite{supp}.
%


\section{Conclusions}\label{conclusions}

We developed a formalism to calculate {\em analytically} all
three-wave mixing processes, sum and difference frequency generation,
second harmonic generation \cite{raksha} and
optical rectification, for slightly deformed thin cylinders with a
simple noncentrosymmetric cross-section. Our theory was developed
in $2D$ within the long-wavelength approximation, assuming
translational symmetry along the axis
of the cylinder, and is of a perturbative nature, assuming the geometry
is controlled by a small deformation parameter.
We first generalized the {\em dipolium} model to calculate the DF response of a
semi-infinite homogeneous media, and used the results to compute
the bulk and surface contributions to the DF polarization of the cylinder,
assuming its surface is smooth and thus locally flat. This
polarization is a source for the near DF
fields from which we identified the total DF electric dipole $\bm p_-$ and
quadrupole $\bm Q_-$, and we also obtained the nonlinear magnetic dipole
$\bm m_-$. We thus obtained all the finite components of the
corresponding hyperpolarizability tensors which were of order one in
the deformation parameter for the electric dipolar case and of order zero in the
other cases. We also calculated the radiation fields, radiation
patterns and conversion efficiency. A simple extension
allowed us to also obtain the induced moments,
hyperpolarizabilities and the radiation corresponding to SF, SH and
OR.

Our results are written in terms of
the linear dielectric response of the system evaluated at the relevant
frequencies.
We illustrated them by calculating and analyzing the SF/DF
hyperpolarizabilities, radiation patterns and
efficiencies for a model harmonic
dielectric, and we interpreted their resonant structure, related to
the excitation of surface plasmon polaritons at the input and/or output
frequencies. We found a strong DFG when $\omega_1\approx
\omega_{\mathrm{spp}}\approx \omega_2$, corresponding to a small but
finite $\omega_{-}$, suggesting that our system might yield an
efficient generation of radiation in the THz regime. We further
identified the regions in the frequency space where different
multipolar contributions became dominant and we illustrated them
through calculations of the corresponding radiation patterns.
The nonlinear
magnetic dipole moment was found to dominate for most frequency
combinations, but it is zero for the SH case. Our
results show that for some
$\omega_1,\omega_2$ frequency combinations,
the electric-dipolar contribution may become comparable to the
electric-quadrupolar or the magnetic-dipolar contributions even for
very slightly deformed cylinders.

In summary, we  developed an analytical formalism that allowed us to
explore all three-wave mixing processes at $2D$ nanocylinders made up
of centrosymmetric materials but with a noncentrosymmetric
geometry. Although
we developed the model for an harmonic {\em dipolium} model, the results are
written in terms of the dielectric function of the material evaluated
at the relevant input and output frequencies. Thus, by substituting
the appropriate response functions, our results may be
applied to arbitrary dielectrics. Furthermore, it may be shown that
the results agree with those of a local {\em jellium} model, so they
may also be applied to metals \cite{hydro}.
Our theory does not take into account effects
related to crystal structure, the presence of surface states, and
surface reconstructions and relaxation. Nevertheless, our results
allow a quantification of the expected efficiency of the different
processes, and, in particular, they show
that electric electric dipolar contributions may dominate the
quadrupolar and magnetic dipolar ones at certain frequency
combinations even for very small deformations. Thus, ordinary
centrosymmetric materials textured with noncentrosymmetric patterns
may provide competitive sources of optical sum and difference
frequency generation for processes such as conversion of light into the THz
regime. Furthermore, our model provides analytical expressions against
which numerical computational schemes may be tested.


\acknowledgments
This work was supported by DGAPA-UNAM under grant IN111119
(WLM) and by CONACyT (RS). We acknowledge useful
discussions with B. S. Mendoza.

\appendix
\section*{Appendix}
\renewcommand{\theequation}{A\thesection.\arabic{equation}}

In this Appendix, we calculate the electromagnetic fields radiated in
$2D$ by a magnetic dipole. A detailed description of the derivation is
not presented here as a similar calculation was discussed in the Appendix of
Ref. \cite{raksha}, where radiation from an electric dipole and a
quadrupole were considered. Eq. (A6) of Ref. \cite{raksha} is an
expression for the
vector potential in $2D$ radiated by a harmonically varying monochromatic
current distribution $\bm{J}(\bm{r},t)$, expressed as a power series
in the diameter of the system. The
second term of this series is
\begin{equation}\label{quad1}
  \bm A^{(1)}(\bm r)=\frac{1}{c}\sqrt{\frac{2\pi}{kr}}
  e^{i\pi/4}e^{ikr}(-ik)\int d^2r'\, \bm J(\bm r')(\hat{\bm r}\cdot \bm r'),
\end{equation}
Its integrand can be written as the sum of a symmetric and an antisymmetric part,
$\bm J(\bm r')(\hat{\bm r}\cdot \bm r')
=(1/2)[\bm J(\bm r')(\hat{\bm r}\cdot \bm r') +
    \bm r'(\hat{\bm r}\cdot\bm J(\bm r'))]
  +(1/2)[\bm J(\bm r')(\hat{\bm r}\cdot \bm r') -
  \bm r'(\hat{\bm r}\cdot \bm J(\bm r'))]$. The former yields the
  electric quadrupolar radiation while latter corresponds to
  the contribution of the magnetic dipole, which can be
  written as
\begin{equation}
  \bm{A}^m(\bm{r})=\sqrt{\frac{2\pi}{kr}}e^{i\pi/4}e^{ikr}(ik)
                    \bm{\hat{r}}\times\bm{m},
  \label{vecpotmd}
\end{equation}
where $\bm{m}$ is the magnetic dipole moment per unit length,
\begin{equation}
  \bm{m}=\frac{1}{2c}\int d^2r'\, \bm{r'}\times\bm{J}(\bm{r'}).
\end{equation}

As mentioned in Ref. \cite{raksha}, we may obtain the corresponding electromagnetic
radiation field as $\bm B^m=ik\hat{\bm
  r}\times\bm A^m$ and $\bm E^m=\bm B^m\times \hat{\bm r}$.

\bibliography{ref}

\begin{thebibliography}{42}%
\makeatletter
\providecommand \@ifxundefined [1]{%
 \@ifx{#1\undefined}
}%
\providecommand \@ifnum [1]{%
 \ifnum #1\expandafter \@firstoftwo
 \else \expandafter \@secondoftwo
 \fi
}%
\providecommand \@ifx [1]{%
 \ifx #1\expandafter \@firstoftwo
 \else \expandafter \@secondoftwo
 \fi
}%
\providecommand \natexlab [1]{#1}%
\providecommand \enquote  [1]{``#1''}%
\providecommand \bibnamefont  [1]{#1}%
\providecommand \bibfnamefont [1]{#1}%
\providecommand \citenamefont [1]{#1}%
\providecommand \href@noop [0]{\@secondoftwo}%
\providecommand \href [0]{\begingroup \@sanitize@url \@href}%
\providecommand \@href[1]{\@@startlink{#1}\@@href}%
\providecommand \@@href[1]{\endgroup#1\@@endlink}%
\providecommand \@sanitize@url [0]{\catcode `\\12\catcode `\$12\catcode
  `\&12\catcode `\#12\catcode `\^12\catcode `\_12\catcode `\%12\relax}%
\providecommand \@@startlink[1]{}%
\providecommand \@@endlink[0]{}%
\providecommand \url  [0]{\begingroup\@sanitize@url \@url }%
\providecommand \@url [1]{\endgroup\@href {#1}{\urlprefix }}%
\providecommand \urlprefix  [0]{URL }%
\providecommand \Eprint [0]{\href }%
\providecommand \doibase [0]{https://doi.org/}%
\providecommand \selectlanguage [0]{\@gobble}%
\providecommand \bibinfo  [0]{\@secondoftwo}%
\providecommand \bibfield  [0]{\@secondoftwo}%
\providecommand \translation [1]{[#1]}%
\providecommand \BibitemOpen [0]{}%
\providecommand \bibitemStop [0]{}%
\providecommand \bibitemNoStop [0]{.\EOS\space}%
\providecommand \EOS [0]{\spacefactor3000\relax}%
\providecommand \BibitemShut  [1]{\csname bibitem#1\endcsname}%
\let\auto@bib@innerbib\@empty
\bibitem [{\citenamefont {Dudley}\ \emph {et~al.}(2006)\citenamefont {Dudley},
  \citenamefont {Genty},\ and\ \citenamefont {Coen}}]{dudley}%
  \BibitemOpen
  \bibfield  {author} {\bibinfo {author} {\bibfnamefont {J.~M.}\ \bibnamefont
  {Dudley}}, \bibinfo {author} {\bibfnamefont {G.}~\bibnamefont {Genty}},\ and\
  \bibinfo {author} {\bibfnamefont {S.}~\bibnamefont {Coen}},\ }\bibfield
  {title} {\bibinfo {title} {Supercontinuum generation in photonic crystal
  fiber},\ }\href {https://doi.org/10.1103/RevModPhys.78.1135} {\bibfield
  {journal} {\bibinfo  {journal} {Rev. Mod. Phys.}\ }\textbf {\bibinfo {volume}
  {78}},\ \bibinfo {pages} {1135} (\bibinfo {year} {2006})}\BibitemShut
  {NoStop}%
\bibitem [{\citenamefont {Yariv}(1978)}]{yariv}%
  \BibitemOpen
  \bibfield  {author} {\bibinfo {author} {\bibfnamefont {A.}~\bibnamefont
  {Yariv}},\ }\bibfield  {title} {\bibinfo {title} {Four wave nonlinear optical
  mixing as real time holography},\ }\href
  {https://doi.org/https://doi.org/10.1016/0030-4018(78)90079-2} {\bibfield
  {journal} {\bibinfo  {journal} {Optics Communications}\ }\textbf {\bibinfo
  {volume} {25}},\ \bibinfo {pages} {23 } (\bibinfo {year} {1978})}\BibitemShut
  {NoStop}%
\bibitem [{\citenamefont {Kroll}(1962)}]{opa}%
  \BibitemOpen
  \bibfield  {author} {\bibinfo {author} {\bibfnamefont {N.~M.}\ \bibnamefont
  {Kroll}},\ }\bibfield  {title} {\bibinfo {title} {Parametric amplification in
  spatially extended media and application to the design of tuneable
  oscillators at optical frequencies},\ }\href
  {https://doi.org/10.1103/PhysRev.127.1207} {\bibfield  {journal} {\bibinfo
  {journal} {Phys. Rev.}\ }\textbf {\bibinfo {volume} {127}},\ \bibinfo {pages}
  {1207} (\bibinfo {year} {1962})}\BibitemShut {NoStop}%
\bibitem [{\citenamefont {Kaiser}()}]{kaiser}%
  \BibitemOpen
  \bibfield  {author} {\bibinfo {author} {\bibfnamefont {W.}~\bibnamefont
  {Kaiser}},\ }\href@noop {} {\emph {\bibinfo {title} {{Ultrashort Laser
  Pulses}}}}\BibitemShut {NoStop}%
\bibitem [{\citenamefont {Park}\ \emph {et~al.}(2011)\citenamefont {Park},
  \citenamefont {Kim}, \citenamefont {Choi}, \citenamefont {Lee}, \citenamefont
  {Kim}, \citenamefont {Kling}, \citenamefont {Stockman},\ and\ \citenamefont
  {Kim}}]{ultra}%
  \BibitemOpen
  \bibfield  {author} {\bibinfo {author} {\bibfnamefont {I.-Y.}\ \bibnamefont
  {Park}}, \bibinfo {author} {\bibfnamefont {S.}~\bibnamefont {Kim}}, \bibinfo
  {author} {\bibfnamefont {J.}~\bibnamefont {Choi}}, \bibinfo {author}
  {\bibfnamefont {D.-H.}\ \bibnamefont {Lee}}, \bibinfo {author} {\bibfnamefont
  {Y.-J.}\ \bibnamefont {Kim}}, \bibinfo {author} {\bibfnamefont {M.~F.}\
  \bibnamefont {Kling}}, \bibinfo {author} {\bibfnamefont {M.~I.}\ \bibnamefont
  {Stockman}},\ and\ \bibinfo {author} {\bibfnamefont {S.-W.}\ \bibnamefont
  {Kim}},\ }\bibfield  {title} {\bibinfo {title} {Plasmonic generation of
  ultrashort extreme-ultraviolet light pulses},\ }\href
  {https://doi.org/10.1038/nphoton.2011.258} {\bibfield  {journal} {\bibinfo
  {journal} {Nature Photonics}\ }\textbf {\bibinfo {volume} {5}},\ \bibinfo
  {pages} {677} (\bibinfo {year} {2011})}\BibitemShut {NoStop}%
\bibitem [{\citenamefont {Monticone}\ and\ \citenamefont
  {Al{\`{u}}}(2017)}]{nanophotonic}%
  \BibitemOpen
  \bibfield  {author} {\bibinfo {author} {\bibfnamefont {F.}~\bibnamefont
  {Monticone}}\ and\ \bibinfo {author} {\bibfnamefont {A.}~\bibnamefont
  {Al{\`{u}}}},\ }\bibfield  {title} {\bibinfo {title} {Metamaterial, plasmonic
  and nanophotonic devices},\ }\href {https://doi.org/10.1088/1361-6633/aa518f}
  {\bibfield  {journal} {\bibinfo  {journal} {Reports on Progress in Physics}\
  }\textbf {\bibinfo {volume} {80}},\ \bibinfo {pages} {036401} (\bibinfo
  {year} {2017})}\BibitemShut {NoStop}%
\bibitem [{\citenamefont {Smirnova}\ and\ \citenamefont
  {Kivshar}(2016)}]{rev1nlo}%
  \BibitemOpen
  \bibfield  {author} {\bibinfo {author} {\bibfnamefont {D.}~\bibnamefont
  {Smirnova}}\ and\ \bibinfo {author} {\bibfnamefont {Y.~S.}\ \bibnamefont
  {Kivshar}},\ }\bibfield  {title} {\bibinfo {title} {Multipolar nonlinear
  nanophotonics},\ }\href {https://doi.org/10.1364/OPTICA.3.001241} {\bibfield
  {journal} {\bibinfo  {journal} {Optica}\ }\textbf {\bibinfo {volume} {3}},\
  \bibinfo {pages} {1241} (\bibinfo {year} {2016})}\BibitemShut {NoStop}%
\bibitem [{\citenamefont {Kauranen}\ and\ \citenamefont
  {Zayats}(2012)}]{kauranen}%
  \BibitemOpen
  \bibfield  {author} {\bibinfo {author} {\bibfnamefont {M.}~\bibnamefont
  {Kauranen}}\ and\ \bibinfo {author} {\bibfnamefont {A.~V.}\ \bibnamefont
  {Zayats}},\ }\bibfield  {title} {\bibinfo {title} {Nonlinear plasmonics},\
  }\href {https://doi.org/10.1038/nphoton.2012.244} {\bibfield  {journal}
  {\bibinfo  {journal} {Nature Photonics}\ }\textbf {\bibinfo {volume} {6}},\
  \bibinfo {pages} {737} (\bibinfo {year} {2012})}\BibitemShut {NoStop}%
\bibitem [{\citenamefont {Czaplicki}\ \emph {et~al.}(2015)\citenamefont
  {Czaplicki}, \citenamefont {Mäkitalo}, \citenamefont {Siikanen},
  \citenamefont {Husu}, \citenamefont {Lehtolahti}, \citenamefont {Kuittinen},\
  and\ \citenamefont {Kauranen}}]{czaplicki}%
  \BibitemOpen
  \bibfield  {author} {\bibinfo {author} {\bibfnamefont {R.}~\bibnamefont
  {Czaplicki}}, \bibinfo {author} {\bibfnamefont {J.}~\bibnamefont
  {Mäkitalo}}, \bibinfo {author} {\bibfnamefont {R.}~\bibnamefont {Siikanen}},
  \bibinfo {author} {\bibfnamefont {H.}~\bibnamefont {Husu}}, \bibinfo {author}
  {\bibfnamefont {J.}~\bibnamefont {Lehtolahti}}, \bibinfo {author}
  {\bibfnamefont {M.}~\bibnamefont {Kuittinen}},\ and\ \bibinfo {author}
  {\bibfnamefont {M.}~\bibnamefont {Kauranen}},\ }\bibfield  {title} {\bibinfo
  {title} {Second-harmonic generation from metal nanoparticles: Resonance
  enhancement versus particle geometry},\ }\href
  {https://doi.org/10.1021/nl503901e} {\bibfield  {journal} {\bibinfo
  {journal} {Nano Letters}\ }\textbf {\bibinfo {volume} {15}},\ \bibinfo
  {pages} {530} (\bibinfo {year} {2015})},\ \bibinfo {note} {pMID: 25521745},\
  \Eprint {https://arxiv.org/abs/https://doi.org/10.1021/nl503901e}
  {https://doi.org/10.1021/nl503901e} \BibitemShut {NoStop}%
\bibitem [{\citenamefont {Liu}\ \emph {et~al.}(2016)\citenamefont {Liu},
  \citenamefont {Sinclair}, \citenamefont {Saravi}, \citenamefont {Keeler},
  \citenamefont {Yang}, \citenamefont {Reno}, \citenamefont {Peake},
  \citenamefont {Setzpfandt}, \citenamefont {Staude}, \citenamefont {Pertsch},\
  and\ \citenamefont {Brener}}]{diel-nl}%
  \BibitemOpen
  \bibfield  {author} {\bibinfo {author} {\bibfnamefont {S.}~\bibnamefont
  {Liu}}, \bibinfo {author} {\bibfnamefont {M.~B.}\ \bibnamefont {Sinclair}},
  \bibinfo {author} {\bibfnamefont {S.}~\bibnamefont {Saravi}}, \bibinfo
  {author} {\bibfnamefont {G.~A.}\ \bibnamefont {Keeler}}, \bibinfo {author}
  {\bibfnamefont {Y.}~\bibnamefont {Yang}}, \bibinfo {author} {\bibfnamefont
  {J.}~\bibnamefont {Reno}}, \bibinfo {author} {\bibfnamefont {G.~M.}\
  \bibnamefont {Peake}}, \bibinfo {author} {\bibfnamefont {F.}~\bibnamefont
  {Setzpfandt}}, \bibinfo {author} {\bibfnamefont {I.}~\bibnamefont {Staude}},
  \bibinfo {author} {\bibfnamefont {T.}~\bibnamefont {Pertsch}},\ and\ \bibinfo
  {author} {\bibfnamefont {I.}~\bibnamefont {Brener}},\ }\bibfield  {title}
  {\bibinfo {title} {Resonantly enhanced second-harmonic generation using
  iii–v semiconductor all-dielectric metasurfaces},\ }\href
  {https://doi.org/10.1021/acs.nanolett.6b01816} {\bibfield  {journal}
  {\bibinfo  {journal} {Nano Letters}\ }\textbf {\bibinfo {volume} {16}},\
  \bibinfo {pages} {5426} (\bibinfo {year} {2016})},\ \bibinfo {note} {pMID:
  27501472},\ \Eprint
  {https://arxiv.org/abs/https://doi.org/10.1021/acs.nanolett.6b01816}
  {https://doi.org/10.1021/acs.nanolett.6b01816} \BibitemShut {NoStop}%
\bibitem [{\citenamefont {Metzger}\ \emph {et~al.}(2014)\citenamefont
  {Metzger}, \citenamefont {Hentschel}, \citenamefont {Schumacher},
  \citenamefont {Lippitz}, \citenamefont {Ye}, \citenamefont {Murray},
  \citenamefont {Knabe}, \citenamefont {Buse},\ and\ \citenamefont
  {Giessen}}]{thg-diel}%
  \BibitemOpen
  \bibfield  {author} {\bibinfo {author} {\bibfnamefont {B.}~\bibnamefont
  {Metzger}}, \bibinfo {author} {\bibfnamefont {M.}~\bibnamefont {Hentschel}},
  \bibinfo {author} {\bibfnamefont {T.}~\bibnamefont {Schumacher}}, \bibinfo
  {author} {\bibfnamefont {M.}~\bibnamefont {Lippitz}}, \bibinfo {author}
  {\bibfnamefont {X.}~\bibnamefont {Ye}}, \bibinfo {author} {\bibfnamefont
  {C.~B.}\ \bibnamefont {Murray}}, \bibinfo {author} {\bibfnamefont
  {B.}~\bibnamefont {Knabe}}, \bibinfo {author} {\bibfnamefont
  {K.}~\bibnamefont {Buse}},\ and\ \bibinfo {author} {\bibfnamefont
  {H.}~\bibnamefont {Giessen}},\ }\bibfield  {title} {\bibinfo {title}
  {Doubling the efficiency of third harmonic generation by positioning ito
  nanocrystals into the hot-spot of plasmonic gap-antennas},\ }\href
  {https://doi.org/10.1021/nl500913t} {\bibfield  {journal} {\bibinfo
  {journal} {Nano Letters}\ }\textbf {\bibinfo {volume} {14}},\ \bibinfo
  {pages} {2867} (\bibinfo {year} {2014})},\ \bibinfo {note} {pMID: 24730433},\
  \Eprint {https://arxiv.org/abs/https://doi.org/10.1021/nl500913t}
  {https://doi.org/10.1021/nl500913t} \BibitemShut {NoStop}%
\bibitem [{\citenamefont {Gili}\ \emph {et~al.}(2016)\citenamefont {Gili},
  \citenamefont {Carletti}, \citenamefont {Locatelli}, \citenamefont {Rocco},
  \citenamefont {Finazzi}, \citenamefont {Ghirardini}, \citenamefont {Favero},
  \citenamefont {Gomez}, \citenamefont {Lema\^{i}tre}, \citenamefont
  {Celebrano}, \citenamefont {Angelis},\ and\ \citenamefont {Leo}}]{gili}%
  \BibitemOpen
  \bibfield  {author} {\bibinfo {author} {\bibfnamefont {V.~F.}\ \bibnamefont
  {Gili}}, \bibinfo {author} {\bibfnamefont {L.}~\bibnamefont {Carletti}},
  \bibinfo {author} {\bibfnamefont {A.}~\bibnamefont {Locatelli}}, \bibinfo
  {author} {\bibfnamefont {D.}~\bibnamefont {Rocco}}, \bibinfo {author}
  {\bibfnamefont {M.}~\bibnamefont {Finazzi}}, \bibinfo {author} {\bibfnamefont
  {L.}~\bibnamefont {Ghirardini}}, \bibinfo {author} {\bibfnamefont
  {I.}~\bibnamefont {Favero}}, \bibinfo {author} {\bibfnamefont
  {C.}~\bibnamefont {Gomez}}, \bibinfo {author} {\bibfnamefont
  {A.}~\bibnamefont {Lema\^{i}tre}}, \bibinfo {author} {\bibfnamefont
  {M.}~\bibnamefont {Celebrano}}, \bibinfo {author} {\bibfnamefont {C.~D.}\
  \bibnamefont {Angelis}},\ and\ \bibinfo {author} {\bibfnamefont
  {G.}~\bibnamefont {Leo}},\ }\bibfield  {title} {\bibinfo {title} {Monolithic
  algaas second-harmonic nanoantennas},\ }\href
  {https://doi.org/10.1364/OE.24.015965} {\bibfield  {journal} {\bibinfo
  {journal} {Opt. Express}\ }\textbf {\bibinfo {volume} {24}},\ \bibinfo
  {pages} {15965} (\bibinfo {year} {2016})}\BibitemShut {NoStop}%
\bibitem [{\citenamefont {Meza}\ \emph {et~al.}(2019)\citenamefont {Meza},
  \citenamefont {Mendoza},\ and\ \citenamefont {Moch\'an}}]{ulises}%
  \BibitemOpen
  \bibfield  {author} {\bibinfo {author} {\bibfnamefont {U.~R.}\ \bibnamefont
  {Meza}}, \bibinfo {author} {\bibfnamefont {B.~S.}\ \bibnamefont {Mendoza}},\
  and\ \bibinfo {author} {\bibfnamefont {W.~L.}\ \bibnamefont {Moch\'an}},\
  }\bibfield  {title} {\bibinfo {title} {Second-harmonic generation in
  nanostructured metamaterials},\ }\href
  {https://doi.org/10.1103/PhysRevB.99.125408} {\bibfield  {journal} {\bibinfo
  {journal} {Phys. Rev. B}\ }\textbf {\bibinfo {volume} {99}},\ \bibinfo
  {pages} {125408} (\bibinfo {year} {2019})}\BibitemShut {NoStop}%
\bibitem [{\citenamefont {Singla}\ and\ \citenamefont
  {Moch\'an}(2019)}]{raksha}%
  \BibitemOpen
  \bibfield  {author} {\bibinfo {author} {\bibfnamefont {R.}~\bibnamefont
  {Singla}}\ and\ \bibinfo {author} {\bibfnamefont {W.~L.}\ \bibnamefont
  {Moch\'an}},\ }\bibfield  {title} {\bibinfo {title} {Analytical theory of
  second harmonic generation from a nanowire with noncentrosymmetric
  geometry},\ }\href {https://doi.org/10.1103/PhysRevB.99.125418} {\bibfield
  {journal} {\bibinfo  {journal} {Phys. Rev. B}\ }\textbf {\bibinfo {volume}
  {99}},\ \bibinfo {pages} {125418} (\bibinfo {year} {2019})}\BibitemShut
  {NoStop}%
\bibitem [{\citenamefont {Wenseleers}\ \emph {et~al.}(2002)\citenamefont
  {Wenseleers}, \citenamefont {Stellacci}, \citenamefont {Meyer-Friedrichsen},
  \citenamefont {Mangel}, \citenamefont {Bauer}, \citenamefont {Pond},
  \citenamefont {Marder},\ and\ \citenamefont {Perry}}]{tpa}%
  \BibitemOpen
  \bibfield  {author} {\bibinfo {author} {\bibfnamefont {W.}~\bibnamefont
  {Wenseleers}}, \bibinfo {author} {\bibfnamefont {F.}~\bibnamefont
  {Stellacci}}, \bibinfo {author} {\bibfnamefont {T.}~\bibnamefont
  {Meyer-Friedrichsen}}, \bibinfo {author} {\bibfnamefont {T.}~\bibnamefont
  {Mangel}}, \bibinfo {author} {\bibfnamefont {C.~A.}\ \bibnamefont {Bauer}},
  \bibinfo {author} {\bibfnamefont {S.~J.~K.}\ \bibnamefont {Pond}}, \bibinfo
  {author} {\bibfnamefont {S.~R.}\ \bibnamefont {Marder}},\ and\ \bibinfo
  {author} {\bibfnamefont {J.~W.}\ \bibnamefont {Perry}},\ }\bibfield  {title}
  {\bibinfo {title} {Five orders-of-magnitude enhancement of two-photon
  absorption for dyes on silver nanoparticle fractal clusters},\ }\href
  {https://doi.org/10.1021/jp014675f} {\bibfield  {journal} {\bibinfo
  {journal} {The Journal of Physical Chemistry B}\ }\textbf {\bibinfo {volume}
  {106}},\ \bibinfo {pages} {6853} (\bibinfo {year} {2002})},\ \Eprint
  {https://arxiv.org/abs/https://doi.org/10.1021/jp014675f}
  {https://doi.org/10.1021/jp014675f} \BibitemShut {NoStop}%
\bibitem [{\citenamefont {Lippitz}\ \emph {et~al.}(2005)\citenamefont
  {Lippitz}, \citenamefont {van Dijk},\ and\ \citenamefont {Orrit}}]{thg1}%
  \BibitemOpen
  \bibfield  {author} {\bibinfo {author} {\bibfnamefont {M.}~\bibnamefont
  {Lippitz}}, \bibinfo {author} {\bibfnamefont {M.~A.}\ \bibnamefont {van
  Dijk}},\ and\ \bibinfo {author} {\bibfnamefont {M.}~\bibnamefont {Orrit}},\
  }\bibfield  {title} {\bibinfo {title} {Third-harmonic generation from single
  gold nanoparticles},\ }\href {https://doi.org/10.1021/nl0502571} {\bibfield
  {journal} {\bibinfo  {journal} {Nano Letters}\ }\textbf {\bibinfo {volume}
  {5}},\ \bibinfo {pages} {799} (\bibinfo {year} {2005})},\ \bibinfo {note}
  {pMID: 15826131},\ \Eprint
  {https://arxiv.org/abs/https://doi.org/10.1021/nl0502571}
  {https://doi.org/10.1021/nl0502571} \BibitemShut {NoStop}%
\bibitem [{\citenamefont {Yariv}\ and\ \citenamefont {Pepper}(1977)}]{fwm}%
  \BibitemOpen
  \bibfield  {author} {\bibinfo {author} {\bibfnamefont {A.}~\bibnamefont
  {Yariv}}\ and\ \bibinfo {author} {\bibfnamefont {D.~M.}\ \bibnamefont
  {Pepper}},\ }\bibfield  {title} {\bibinfo {title} {Amplified reflection,
  phase conjugation, and oscillation in degenerate four-wave mixing},\ }\href
  {https://doi.org/10.1364/OL.1.000016} {\bibfield  {journal} {\bibinfo
  {journal} {Opt. Lett.}\ }\textbf {\bibinfo {volume} {1}},\ \bibinfo {pages}
  {16} (\bibinfo {year} {1977})}\BibitemShut {NoStop}%
\bibitem [{\citenamefont {Boyer}\ \emph {et~al.}(2008)\citenamefont {Boyer},
  \citenamefont {Marino}, \citenamefont {Pooser},\ and\ \citenamefont
  {Lett}}]{fwm1}%
  \BibitemOpen
  \bibfield  {author} {\bibinfo {author} {\bibfnamefont {V.}~\bibnamefont
  {Boyer}}, \bibinfo {author} {\bibfnamefont {A.~M.}\ \bibnamefont {Marino}},
  \bibinfo {author} {\bibfnamefont {R.~C.}\ \bibnamefont {Pooser}},\ and\
  \bibinfo {author} {\bibfnamefont {P.~D.}\ \bibnamefont {Lett}},\ }\bibfield
  {title} {\bibinfo {title} {Entangled images from four-wave mixing},\ }\href
  {https://doi.org/10.1126/science.1158275} {\bibfield  {journal} {\bibinfo
  {journal} {Science}\ }\textbf {\bibinfo {volume} {321}},\ \bibinfo {pages}
  {544} (\bibinfo {year} {2008})},\ \Eprint
  {https://arxiv.org/abs/https://science.sciencemag.org/content/321/5888/544.full.pdf}
  {https://science.sciencemag.org/content/321/5888/544.full.pdf} \BibitemShut
  {NoStop}%
\bibitem [{\citenamefont {Ji}\ \emph {et~al.}(2006)\citenamefont {Ji},
  \citenamefont {Zhang}, \citenamefont {Yang},\ and\ \citenamefont
  {Shen}}]{sfgimg1}%
  \BibitemOpen
  \bibfield  {author} {\bibinfo {author} {\bibfnamefont {N.}~\bibnamefont
  {Ji}}, \bibinfo {author} {\bibfnamefont {K.}~\bibnamefont {Zhang}}, \bibinfo
  {author} {\bibfnamefont {H.}~\bibnamefont {Yang}},\ and\ \bibinfo {author}
  {\bibfnamefont {Y.-R.}\ \bibnamefont {Shen}},\ }\bibfield  {title} {\bibinfo
  {title} {Three-dimensional chiral imaging by sum-frequency generation},\
  }\href {https://doi.org/10.1021/ja057775y} {\bibfield  {journal} {\bibinfo
  {journal} {Journal of the American Chemical Society}\ }\textbf {\bibinfo
  {volume} {128}},\ \bibinfo {pages} {3482} (\bibinfo {year} {2006})},\
  \bibinfo {note} {pMID: 16536497},\ \Eprint
  {https://arxiv.org/abs/https://doi.org/10.1021/ja057775y}
  {https://doi.org/10.1021/ja057775y} \BibitemShut {NoStop}%
\bibitem [{\citenamefont {Pikalov}\ \emph {et~al.}(2019)\citenamefont
  {Pikalov}, \citenamefont {Ngo}, \citenamefont {Lee}, \citenamefont {Lee},\
  and\ \citenamefont {Baldelli}}]{sfgimg2}%
  \BibitemOpen
  \bibfield  {author} {\bibinfo {author} {\bibfnamefont {A.~A.}\ \bibnamefont
  {Pikalov}}, \bibinfo {author} {\bibfnamefont {D.}~\bibnamefont {Ngo}},
  \bibinfo {author} {\bibfnamefont {H.~J.}\ \bibnamefont {Lee}}, \bibinfo
  {author} {\bibfnamefont {T.~R.}\ \bibnamefont {Lee}},\ and\ \bibinfo {author}
  {\bibfnamefont {S.}~\bibnamefont {Baldelli}},\ }\bibfield  {title} {\bibinfo
  {title} {Sum frequency generation imaging microscopy of self-assembled
  monolayers on metal surfaces: Factor analysis of mixed monolayers},\ }\href
  {https://doi.org/10.1021/acs.analchem.8b01840} {\bibfield  {journal}
  {\bibinfo  {journal} {Analytical Chemistry}\ }\textbf {\bibinfo {volume}
  {91}},\ \bibinfo {pages} {1269} (\bibinfo {year} {2019})},\ \Eprint
  {https://arxiv.org/abs/https://doi.org/10.1021/acs.analchem.8b01840}
  {https://doi.org/10.1021/acs.analchem.8b01840} \BibitemShut {NoStop}%
\bibitem [{\citenamefont {Moreaux}\ \emph {et~al.}(2000)\citenamefont
  {Moreaux}, \citenamefont {Sandre},\ and\ \citenamefont {Mertz}}]{shgimg1}%
  \BibitemOpen
  \bibfield  {author} {\bibinfo {author} {\bibfnamefont {L.}~\bibnamefont
  {Moreaux}}, \bibinfo {author} {\bibfnamefont {O.}~\bibnamefont {Sandre}},\
  and\ \bibinfo {author} {\bibfnamefont {J.}~\bibnamefont {Mertz}},\ }\bibfield
   {title} {\bibinfo {title} {Membrane imaging by second-harmonic generation
  microscopy},\ }\href {https://doi.org/10.1364/JOSAB.17.001685} {\bibfield
  {journal} {\bibinfo  {journal} {J. Opt. Soc. Am. B}\ }\textbf {\bibinfo
  {volume} {17}},\ \bibinfo {pages} {1685} (\bibinfo {year}
  {2000})}\BibitemShut {NoStop}%
\bibitem [{\citenamefont {Dempsey}\ \emph {et~al.}(2012)\citenamefont
  {Dempsey}, \citenamefont {Fraser},\ and\ \citenamefont {Pantazis}}]{shgimg2}%
  \BibitemOpen
  \bibfield  {author} {\bibinfo {author} {\bibfnamefont {W.~P.}\ \bibnamefont
  {Dempsey}}, \bibinfo {author} {\bibfnamefont {S.~E.}\ \bibnamefont
  {Fraser}},\ and\ \bibinfo {author} {\bibfnamefont {P.}~\bibnamefont
  {Pantazis}},\ }\bibfield  {title} {\bibinfo {title} {Shg nanoprobes:
  Advancing harmonic imaging in biology},\ }\href
  {https://doi.org/10.1002/bies.201100106} {\bibfield  {journal} {\bibinfo
  {journal} {BioEssays}\ }\textbf {\bibinfo {volume} {34}},\ \bibinfo {pages}
  {351} (\bibinfo {year} {2012})},\ \Eprint
  {https://arxiv.org/abs/https://onlinelibrary.wiley.com/doi/pdf/10.1002/bies.201100106}
  {https://onlinelibrary.wiley.com/doi/pdf/10.1002/bies.201100106} \BibitemShut
  {NoStop}%
\bibitem [{\citenamefont {{Scheps}}\ and\ \citenamefont
  {{Myers}}(1994)}]{sfggreen}%
  \BibitemOpen
  \bibfield  {author} {\bibinfo {author} {\bibfnamefont {R.}~\bibnamefont
  {{Scheps}}}\ and\ \bibinfo {author} {\bibfnamefont {J.~F.}\ \bibnamefont
  {{Myers}}},\ }\bibfield  {title} {\bibinfo {title} {Dual-wavelength
  coupled-cavity ti:sapphire laser with active mirror for enhanced red
  operation and efficient intracavity sum frequency generation at 459 nm},\
  }\href {https://doi.org/10.1109/3.291375} {\bibfield  {journal} {\bibinfo
  {journal} {IEEE Journal of Quantum Electronics}\ }\textbf {\bibinfo {volume}
  {30}},\ \bibinfo {pages} {1050} (\bibinfo {year} {1994})}\BibitemShut
  {NoStop}%
\bibitem [{\citenamefont {Lü}\ \emph {et~al.}(2010)\citenamefont {Lü},
  \citenamefont {Zhang}, \citenamefont {Fu}, \citenamefont {Xia}, \citenamefont
  {Zheng},\ and\ \citenamefont {Chen}}]{sfg-yellow}%
  \BibitemOpen
  \bibfield  {author} {\bibinfo {author} {\bibfnamefont {Y.}~\bibnamefont
  {Lü}}, \bibinfo {author} {\bibfnamefont {X.}~\bibnamefont {Zhang}}, \bibinfo
  {author} {\bibfnamefont {X.}~\bibnamefont {Fu}}, \bibinfo {author}
  {\bibfnamefont {J.}~\bibnamefont {Xia}}, \bibinfo {author} {\bibfnamefont
  {T.}~\bibnamefont {Zheng}},\ and\ \bibinfo {author} {\bibfnamefont
  {J.}~\bibnamefont {Chen}},\ }\bibfield  {title} {\bibinfo {title}
  {Diode-pumped nd:{LuVO}4 and nd:{YAG} crystals yellow laser at 594 nm based
  on intracavity sum-frequency generation},\ }\href
  {https://doi.org/10.1002/lapl.201010042} {\bibfield  {journal} {\bibinfo
  {journal} {Laser Physics Letters}\ }\textbf {\bibinfo {volume} {7}},\
  \bibinfo {pages} {634} (\bibinfo {year} {2010})}\BibitemShut {NoStop}%
\bibitem [{\citenamefont {Fradkin}\ \emph {et~al.}(1999)\citenamefont
  {Fradkin}, \citenamefont {Arie}, \citenamefont {Skliar},\ and\ \citenamefont
  {Rosenman}}]{dfg1mir}%
  \BibitemOpen
  \bibfield  {author} {\bibinfo {author} {\bibfnamefont {K.}~\bibnamefont
  {Fradkin}}, \bibinfo {author} {\bibfnamefont {A.}~\bibnamefont {Arie}},
  \bibinfo {author} {\bibfnamefont {A.}~\bibnamefont {Skliar}},\ and\ \bibinfo
  {author} {\bibfnamefont {G.}~\bibnamefont {Rosenman}},\ }\bibfield  {title}
  {\bibinfo {title} {Tunable midinfrared source by difference frequency
  generation in bulk periodically poled ktiopo4},\ }\href
  {https://doi.org/10.1063/1.123408} {\bibfield  {journal} {\bibinfo  {journal}
  {Applied Physics Letters}\ }\textbf {\bibinfo {volume} {74}},\ \bibinfo
  {pages} {914} (\bibinfo {year} {1999})},\ \Eprint
  {https://arxiv.org/abs/https://doi.org/10.1063/1.123408}
  {https://doi.org/10.1063/1.123408} \BibitemShut {NoStop}%
\bibitem [{\citenamefont {Yamaguchi}\ \emph {et~al.}(2018)\citenamefont
  {Yamaguchi}, \citenamefont {Hida}, \citenamefont {Suzuki}, \citenamefont
  {Isa}, \citenamefont {Yoshikiyo}, \citenamefont {Fujii}, \citenamefont
  {Nemoto},\ and\ \citenamefont {Kannari}}]{dfg2mir}%
  \BibitemOpen
  \bibfield  {author} {\bibinfo {author} {\bibfnamefont {Y.}~\bibnamefont
  {Yamaguchi}}, \bibinfo {author} {\bibfnamefont {R.}~\bibnamefont {Hida}},
  \bibinfo {author} {\bibfnamefont {T.}~\bibnamefont {Suzuki}}, \bibinfo
  {author} {\bibfnamefont {F.}~\bibnamefont {Isa}}, \bibinfo {author}
  {\bibfnamefont {K.}~\bibnamefont {Yoshikiyo}}, \bibinfo {author}
  {\bibfnamefont {L.}~\bibnamefont {Fujii}}, \bibinfo {author} {\bibfnamefont
  {H.}~\bibnamefont {Nemoto}},\ and\ \bibinfo {author} {\bibfnamefont
  {F.}~\bibnamefont {Kannari}},\ }\bibfield  {title} {\bibinfo {title} {Shaping
  and amplification of wavelength-tunable mid-infrared femtosecond pulses
  generated by intra-pulse difference-frequency mixing with spectral
  focusing},\ }\href {https://doi.org/10.1364/JOSAB.35.0000C1} {\bibfield
  {journal} {\bibinfo  {journal} {J. Opt. Soc. Am. B}\ }\textbf {\bibinfo
  {volume} {35}},\ \bibinfo {pages} {C1} (\bibinfo {year} {2018})}\BibitemShut
  {NoStop}%
\bibitem [{\citenamefont {Canarelli}\ \emph {et~al.}(1992)\citenamefont
  {Canarelli}, \citenamefont {Benko}, \citenamefont {Curl},\ and\ \citenamefont
  {Tittel}}]{dfg3mir}%
  \BibitemOpen
  \bibfield  {author} {\bibinfo {author} {\bibfnamefont {P.}~\bibnamefont
  {Canarelli}}, \bibinfo {author} {\bibfnamefont {Z.}~\bibnamefont {Benko}},
  \bibinfo {author} {\bibfnamefont {R.}~\bibnamefont {Curl}},\ and\ \bibinfo
  {author} {\bibfnamefont {F.~K.}\ \bibnamefont {Tittel}},\ }\bibfield  {title}
  {\bibinfo {title} {Continuous-wave infrared laser spectrometer based on
  difference frequency generation in aggas2 for high-resolution spectroscopy},\
  }\href {https://doi.org/10.1364/JOSAB.9.000197} {\bibfield  {journal}
  {\bibinfo  {journal} {J. Opt. Soc. Am. B}\ }\textbf {\bibinfo {volume} {9}},\
  \bibinfo {pages} {197} (\bibinfo {year} {1992})}\BibitemShut {NoStop}%
\bibitem [{\citenamefont {Belkin}\ \emph {et~al.}(2008)\citenamefont {Belkin},
  \citenamefont {Capasso}, \citenamefont {Xie}, \citenamefont {Belyanin},
  \citenamefont {Fischer}, \citenamefont {Wittmann},\ and\ \citenamefont
  {Faist}}]{thzdfg1}%
  \BibitemOpen
  \bibfield  {author} {\bibinfo {author} {\bibfnamefont {M.~A.}\ \bibnamefont
  {Belkin}}, \bibinfo {author} {\bibfnamefont {F.}~\bibnamefont {Capasso}},
  \bibinfo {author} {\bibfnamefont {F.}~\bibnamefont {Xie}}, \bibinfo {author}
  {\bibfnamefont {A.}~\bibnamefont {Belyanin}}, \bibinfo {author}
  {\bibfnamefont {M.}~\bibnamefont {Fischer}}, \bibinfo {author} {\bibfnamefont
  {A.}~\bibnamefont {Wittmann}},\ and\ \bibinfo {author} {\bibfnamefont
  {J.}~\bibnamefont {Faist}},\ }\bibfield  {title} {\bibinfo {title} {Room
  temperature terahertz quantum cascade laser source based on intracavity
  difference-frequency generation},\ }\href {https://doi.org/10.1063/1.2919051}
  {\bibfield  {journal} {\bibinfo  {journal} {Applied Physics Letters}\
  }\textbf {\bibinfo {volume} {92}},\ \bibinfo {pages} {201101} (\bibinfo
  {year} {2008})},\ \Eprint
  {https://arxiv.org/abs/https://doi.org/10.1063/1.2919051}
  {https://doi.org/10.1063/1.2919051} \BibitemShut {NoStop}%
\bibitem [{\citenamefont {Vijayraghavan}\ \emph {et~al.}(2012)\citenamefont
  {Vijayraghavan}, \citenamefont {Adams}, \citenamefont {Vizbaras},
  \citenamefont {Jang}, \citenamefont {Grasse}, \citenamefont {Boehm},
  \citenamefont {Amann},\ and\ \citenamefont {Belkin}}]{thzdfg2}%
  \BibitemOpen
  \bibfield  {author} {\bibinfo {author} {\bibfnamefont {K.}~\bibnamefont
  {Vijayraghavan}}, \bibinfo {author} {\bibfnamefont {R.~W.}\ \bibnamefont
  {Adams}}, \bibinfo {author} {\bibfnamefont {A.}~\bibnamefont {Vizbaras}},
  \bibinfo {author} {\bibfnamefont {M.}~\bibnamefont {Jang}}, \bibinfo {author}
  {\bibfnamefont {C.}~\bibnamefont {Grasse}}, \bibinfo {author} {\bibfnamefont
  {G.}~\bibnamefont {Boehm}}, \bibinfo {author} {\bibfnamefont {M.~C.}\
  \bibnamefont {Amann}},\ and\ \bibinfo {author} {\bibfnamefont {M.~A.}\
  \bibnamefont {Belkin}},\ }\bibfield  {title} {\bibinfo {title} {Terahertz
  sources based on Čerenkov difference-frequency generation in quantum cascade
  lasers},\ }\href {https://doi.org/10.1063/1.4729042} {\bibfield  {journal}
  {\bibinfo  {journal} {Applied Physics Letters}\ }\textbf {\bibinfo {volume}
  {100}},\ \bibinfo {pages} {251104} (\bibinfo {year} {2012})},\ \Eprint
  {https://arxiv.org/abs/https://doi.org/10.1063/1.4729042}
  {https://doi.org/10.1063/1.4729042} \BibitemShut {NoStop}%
\bibitem [{\citenamefont {Bachelier}\ \emph {et~al.}(2008)\citenamefont
  {Bachelier}, \citenamefont {Russier-Antoine}, \citenamefont {Benichou},
  \citenamefont {Jonin},\ and\ \citenamefont {Brevet}}]{bachelier}%
  \BibitemOpen
  \bibfield  {author} {\bibinfo {author} {\bibfnamefont {G.}~\bibnamefont
  {Bachelier}}, \bibinfo {author} {\bibfnamefont {I.}~\bibnamefont
  {Russier-Antoine}}, \bibinfo {author} {\bibfnamefont {E.}~\bibnamefont
  {Benichou}}, \bibinfo {author} {\bibfnamefont {C.}~\bibnamefont {Jonin}},\
  and\ \bibinfo {author} {\bibfnamefont {P.-F.}\ \bibnamefont {Brevet}},\
  }\bibfield  {title} {\bibinfo {title} {Multipolar second-harmonic generation
  in noble metal nanoparticles},\ }\href
  {https://doi.org/10.1364/JOSAB.25.000955} {\bibfield  {journal} {\bibinfo
  {journal} {J. Opt. Soc. Am. B}\ }\textbf {\bibinfo {volume} {25}},\ \bibinfo
  {pages} {955} (\bibinfo {year} {2008})}\BibitemShut {NoStop}%
\bibitem [{\citenamefont {Zhou}\ \emph {et~al.}(2010)\citenamefont {Zhou},
  \citenamefont {Lu}, \citenamefont {Liu}, \citenamefont {Gong},\ and\
  \citenamefont {Mao}}]{zhou}%
  \BibitemOpen
  \bibfield  {author} {\bibinfo {author} {\bibfnamefont {R.}~\bibnamefont
  {Zhou}}, \bibinfo {author} {\bibfnamefont {H.}~\bibnamefont {Lu}}, \bibinfo
  {author} {\bibfnamefont {X.}~\bibnamefont {Liu}}, \bibinfo {author}
  {\bibfnamefont {Y.}~\bibnamefont {Gong}},\ and\ \bibinfo {author}
  {\bibfnamefont {D.}~\bibnamefont {Mao}},\ }\bibfield  {title} {\bibinfo
  {title} {Second-harmonic generation from a periodic array of
  noncentrosymmetric nanoholes},\ }\href
  {https://doi.org/10.1364/JOSAB.27.002405} {\bibfield  {journal} {\bibinfo
  {journal} {J. Opt. Soc. Am. B}\ }\textbf {\bibinfo {volume} {27}},\ \bibinfo
  {pages} {2405} (\bibinfo {year} {2010})}\BibitemShut {NoStop}%
\bibitem [{\citenamefont {Luca}\ and\ \citenamefont
  {Cirac\'{i}}(2019)}]{ciraci-dfg}%
  \BibitemOpen
  \bibfield  {author} {\bibinfo {author} {\bibfnamefont {F.~D.}\ \bibnamefont
  {Luca}}\ and\ \bibinfo {author} {\bibfnamefont {C.}~\bibnamefont
  {Cirac\'{i}}},\ }\bibfield  {title} {\bibinfo {title} {Difference-frequency
  generation in plasmonic nanostructures: a parameter-free hydrodynamic
  description},\ }\href {https://doi.org/10.1364/JOSAB.36.001979} {\bibfield
  {journal} {\bibinfo  {journal} {J. Opt. Soc. Am. B}\ }\textbf {\bibinfo
  {volume} {36}},\ \bibinfo {pages} {1979} (\bibinfo {year}
  {2019})}\BibitemShut {NoStop}%
\bibitem [{\citenamefont {{Fang}}\ \emph {et~al.}(2017)\citenamefont {{Fang}},
  \citenamefont {{Huang}}, \citenamefont {{Sha}},\ and\ \citenamefont
  {{Wu}}}]{fdtddf}%
  \BibitemOpen
  \bibfield  {author} {\bibinfo {author} {\bibfnamefont {M.}~\bibnamefont
  {{Fang}}}, \bibinfo {author} {\bibfnamefont {Z.}~\bibnamefont {{Huang}}},
  \bibinfo {author} {\bibfnamefont {W.~E.~I.}\ \bibnamefont {{Sha}}},\ and\
  \bibinfo {author} {\bibfnamefont {X.}~\bibnamefont {{Wu}}},\ }\bibfield
  {title} {\bibinfo {title} {Maxwell–hydrodynamic model for simulating
  nonlinear terahertz generation from plasmonic metasurfaces},\ }\href@noop {}
  {\bibfield  {journal} {\bibinfo  {journal} {IEEE Journal on Multiscale and
  Multiphysics Computational Techniques}\ }\textbf {\bibinfo {volume} {2}},\
  \bibinfo {pages} {194} (\bibinfo {year} {2017})}\BibitemShut {NoStop}%
\bibitem [{\citenamefont {Mendoza}\ and\ \citenamefont
  {Moch\'an}(1996)}]{dipolium}%
  \BibitemOpen
  \bibfield  {author} {\bibinfo {author} {\bibfnamefont {B.~S.}\ \bibnamefont
  {Mendoza}}\ and\ \bibinfo {author} {\bibfnamefont {W.~L.}\ \bibnamefont
  {Moch\'an}},\ }\bibfield  {title} {\bibinfo {title} {Exactly solvable model
  of surface second-harmonic generation},\ }\href
  {https://doi.org/10.1103/PhysRevB.53.4999} {\bibfield  {journal} {\bibinfo
  {journal} {Phys. Rev. B}\ }\textbf {\bibinfo {volume} {53}},\ \bibinfo
  {pages} {4999} (\bibinfo {year} {1996})}\BibitemShut {NoStop}%
\bibitem [{\citenamefont {Maytorena}\ \emph {et~al.}(1998)\citenamefont
  {Maytorena}, \citenamefont {Mendoza},\ and\ \citenamefont
  {Moch\'an}}]{hydro}%
  \BibitemOpen
  \bibfield  {author} {\bibinfo {author} {\bibfnamefont {J.~A.}\ \bibnamefont
  {Maytorena}}, \bibinfo {author} {\bibfnamefont {B.~S.}\ \bibnamefont
  {Mendoza}},\ and\ \bibinfo {author} {\bibfnamefont {W.~L.}\ \bibnamefont
  {Moch\'an}},\ }\bibfield  {title} {\bibinfo {title} {Theory of surface sum
  frequency generation spectroscopy},\ }\href
  {https://doi.org/10.1103/PhysRevB.57.2569} {\bibfield  {journal} {\bibinfo
  {journal} {Phys. Rev. B}\ }\textbf {\bibinfo {volume} {57}},\ \bibinfo
  {pages} {2569} (\bibinfo {year} {1998})}\BibitemShut {NoStop}%
\bibitem [{\citenamefont {Rudnick}\ and\ \citenamefont
  {Stern}(1971)}]{rudnick}%
  \BibitemOpen
  \bibfield  {author} {\bibinfo {author} {\bibfnamefont {J.}~\bibnamefont
  {Rudnick}}\ and\ \bibinfo {author} {\bibfnamefont {E.~A.}\ \bibnamefont
  {Stern}},\ }\bibfield  {title} {\bibinfo {title} {Second-harmonic radiation
  from metal surfaces},\ }\href {https://doi.org/10.1103/PhysRevB.4.4274}
  {\bibfield  {journal} {\bibinfo  {journal} {Phys. Rev. B}\ }\textbf {\bibinfo
  {volume} {4}},\ \bibinfo {pages} {4274} (\bibinfo {year} {1971})}\BibitemShut
  {NoStop}%
\bibitem [{\citenamefont {Recamier}\ \emph {et~al.}(2004)\citenamefont
  {Recamier}, \citenamefont {Mochan},\ and\ \citenamefont
  {Maytorena}}]{recamier}%
  \BibitemOpen
  \bibfield  {author} {\bibinfo {author} {\bibfnamefont {J.}~\bibnamefont
  {Recamier}}, \bibinfo {author} {\bibfnamefont {W.~L.}\ \bibnamefont
  {Mochan}},\ and\ \bibinfo {author} {\bibfnamefont {J.~A.}\ \bibnamefont
  {Maytorena}},\ }\bibfield  {title} {\bibinfo {title} {Exact nonlinear
  response of a harmonic oscillator},\ }in\ \href
  {https://doi.org/10.1117/12.590738} {\emph {\bibinfo {booktitle} {5th
  Iberoamerican Meeting on Optics and 8th Latin American Meeting on Optics,
  Lasers, and Their Applications}}},\ Vol.\ \bibinfo {volume} {5622},\ \bibinfo
  {editor} {edited by\ \bibinfo {editor} {\bibfnamefont {A.~M.}\ \bibnamefont
  {O.}}\ and\ \bibinfo {editor} {\bibfnamefont {J.~L.}\ \bibnamefont {Paz}}},\
  \bibinfo {organization} {International Society for Optics and Photonics}\
  (\bibinfo  {publisher} {SPIE},\ \bibinfo {year} {2004})\ pp.\ \bibinfo
  {pages} {513 -- 517}\BibitemShut {NoStop}%
\bibitem [{\citenamefont {Jackson}(1975)}]{jackson}%
  \BibitemOpen
  \bibfield  {author} {\bibinfo {author} {\bibfnamefont {J.~D.}\ \bibnamefont
  {Jackson}},\ }\href {https://cds.cern.ch/record/100964} {\emph {\bibinfo
  {title} {{Classical electrodynamics; 2nd ed.}}}}\ (\bibinfo  {publisher}
  {Wiley},\ \bibinfo {address} {New York, NY},\ \bibinfo {year}
  {1975})\BibitemShut {NoStop}%
\bibitem [{\citenamefont {Agranovich}\ and\ \citenamefont
  {Gartstein}(2009)}]{agranovich1}%
  \BibitemOpen
  \bibfield  {author} {\bibinfo {author} {\bibfnamefont {V.}~\bibnamefont
  {Agranovich}}\ and\ \bibinfo {author} {\bibfnamefont {Y.}~\bibnamefont
  {Gartstein}},\ }\bibfield  {title} {\bibinfo {title} {Electrodynamics of
  metamaterials and the landau–lifshitz approach to the magnetic
  permeability},\ }\href
  {https://doi.org/https://doi.org/10.1016/j.metmat.2009.02.002} {\bibfield
  {journal} {\bibinfo  {journal} {Metamaterials}\ }\textbf {\bibinfo {volume}
  {3}},\ \bibinfo {pages} {1 } (\bibinfo {year} {2009})}\BibitemShut {NoStop}%
\bibitem [{\citenamefont {Ashcroft}\ and\ \citenamefont
  {Mermin}(1976)}]{ashcroft}%
  \BibitemOpen
  \bibfield  {author} {\bibinfo {author} {\bibfnamefont {N.}~\bibnamefont
  {Ashcroft}}\ and\ \bibinfo {author} {\bibfnamefont {N.}~\bibnamefont
  {Mermin}},\ }\href@noop {} {\emph {\bibinfo {title} {{Solid State
  Physics}}}}\ (\bibinfo  {publisher} {Saunders College},\ \bibinfo {address}
  {Philadelphia},\ \bibinfo {year} {1976})\BibitemShut {NoStop}%
\bibitem [{\citenamefont {Petukhov}\ \emph {et~al.}(1998)\citenamefont
  {Petukhov}, \citenamefont {Brudny}, \citenamefont {Moch\'an}, \citenamefont
  {Maytorena}, \citenamefont {Mendoza},\ and\ \citenamefont
  {Rasing}}]{petukhov3}%
  \BibitemOpen
  \bibfield  {author} {\bibinfo {author} {\bibfnamefont {A.~V.}\ \bibnamefont
  {Petukhov}}, \bibinfo {author} {\bibfnamefont {V.~L.}\ \bibnamefont
  {Brudny}}, \bibinfo {author} {\bibfnamefont {W.~L.}\ \bibnamefont
  {Moch\'an}}, \bibinfo {author} {\bibfnamefont {J.~A.}\ \bibnamefont
  {Maytorena}}, \bibinfo {author} {\bibfnamefont {B.~S.}\ \bibnamefont
  {Mendoza}},\ and\ \bibinfo {author} {\bibfnamefont {T.}~\bibnamefont
  {Rasing}},\ }\bibfield  {title} {\bibinfo {title} {Energy conservation and
  the manley-rowe relations in surface nonlinear-optical spectroscopy},\ }\href
  {https://doi.org/10.1103/PhysRevLett.81.566} {\bibfield  {journal} {\bibinfo
  {journal} {Phys. Rev. Lett.}\ }\textbf {\bibinfo {volume} {81}},\ \bibinfo
  {pages} {566} (\bibinfo {year} {1998})}\BibitemShut {NoStop}%
\bibitem [{sup()}]{supp}%
  \BibitemOpen
  \href@noop {} {\bibinfo {title} {Supplementary video: An animation
  illustrating the evolution of the nonlinear electromagnetic radiation pattern
  for different input frequencies. the radiation pattern are plotted
  superimposed on the frequency space map which shows the region of the largest
  contributing parameter to the nonlinear efficiency. a square point moving
  across the map denotes the values of the two input frequencies to which the
  pattern corresponds.}}\BibitemShut {Stop}%
\end{thebibliography}%
\end{document}